\documentclass[11pt,twocolumn]{article}
\pdfoutput=1
\usepackage[utf8]{inputenc}
\usepackage[left=1cm, right=1.5cm, top=2cm, bottom=2cm]{geometry}

\usepackage{graphicx} 
\usepackage[T1]{fontenc}
\usepackage{lmodern}
\usepackage[english]{babel}
\usepackage[autostyle]{csquotes}
\usepackage[]{physics}
\usepackage{amsmath}
\usepackage{xcolor}
\usepackage[version=3]{mhchem}
\usepackage{lscape}
\usepackage{caption}
\usepackage{subfig}
\usepackage{siunitx}
\usepackage{booktabs}
\usepackage{multirow}
\usepackage{multicol,cleveref}
\usepackage{tabularx}
\usepackage[normalem]{ulem}
\useunder{\uline}{\ul}{}

\newcommand{\X}{$X {}^2\Sigma^+$}
\newcommand{\A}{$A {}^2\Pi$}

\newcommand{\cm}{cm$^{-1}$}

\title{Diatomic rovibronic transitions as potential probes for proton-to-electron mass ratio across cosmological time}
\author{Anna-Maree Syme,$^{1}$ Adam Mousley,$^{2}$ Maria Cunningham,$^{3}$ and Laura K McKemmish$^{1}$*
\\
\\
$^{1}$School of Chemistry, University of New South Wales, 2052 Sydney, Australia\\
$^{2}$Department of Physics, Aberystwyth University,\\ Penglais, Aberystwyth, Ceredigion,
SY23 3BZ, United Kingdom\\
$^{3}$School of Physics, University of New South Wales, 2052 Sydney, Australia \\
* l.mckemmish@unsw.edu.au
}

\date{December 2019}

\begin{document}
\onecolumn
\maketitle

\begin{abstract}
    Astrophysical molecular spectroscopy is an important method of searching for new physics through probing the variation of the proton-to-electron mass ratio, $\mu$, with existing constraints limiting variation to a fractional change of less than 10$^{-17}$/year. To improve on this constraint and therefore provide better guidance to theories of new physics, new molecular probes will be useful. These probes must have spectral transitions that are observable astrophysically and have different sensitivities to variation in the proton-to-electron mass ratio. Here, we concisely detail how astrophysical observations constrain the set of potential molecular probes and promising sensitive transitions based on how the frequency and intensity of these transitions align with available telescopes and observational constraints. Our detailed investigation focuses on rovibronic transitions in astrophysical diatomic molecules, using the spectroscopic models of 11 diatomics to identify sensitive transitions and probe how they generally arise in real complex molecules with many electronic states and fine structure. While none of the 11 diatomics investigated have sensitive transitions likely to be astrophysically observable, we have found that at high temperatures (1000 K) five of these diatomics have a significant number of low intensity sensitive transitions arising from an accidental near-degeneracy between vibrational levels in the ground and excited electronic state. This insight enables screening of all astrophysical diatomics as potential probes of proton-to-electron mass variation, with CN, CP, SiN and SiC being the most promising candidates for further investigation for sensitivity in rovibronic transitions. 
    
    \vspace{4em}
    
\end{abstract}

\twocolumn
\section{Introduction}


Though general relativity and the Standard Model have been outstandingly successful theories of the physics of the heavy and the small respectively, our understanding of physics is incomplete due to the incompatibility of these two theories\cite{08Al.mp2me,17Ra.mp2me} as well as major unexplained observations. Most startlingly, perhaps, is the fact that only 5\% of the universe is made up of matter that we currently understand; the remainder is composed of dark matter and dark energy which we cannot yet explain though we can clearly observe their effects. \cite{08MuFlMu.mp2me}  
Solving these mysteries in a major ongoing area of research in modern physics that has led to a large number of competing possible theories (see Uzan 2011 \cite{11Uz.mp2me} and references within for an introduction to some of these theories). Many of these theories allow for or require the variation of one or more fundamental constants, \cite{18SaBuDe.mp2me} either spatially or temporally. Thus, high accuracy measurements of the variation in fundamental constants is a sensitive test of potential theories of modern physics.  \cite{15Uz.mp2me}

The two dimensionless fundamental constants that are most often probed astrophysically are the fine structure constant, $\alpha$, and the proton-to-electron mass ration, $\mu$. 
\cite{14JaBeUb.mp2me, 16UbBaSa.mp2me, 11Th.mp2me, 18SaBuDe.mp2me,15Uz.mp2me, 17Ma.mp2me} Since it is of course impossible to directly measure $\alpha$ or $\mu$ across cosmological time, we usually rely on observing physical properties that are affected by a change in these fundamental constants. Variation in $\alpha$ is usually observed through atomic lines. \cite{99WeFlCh.mp2me} 
To search for variation in $\mu$, we observe the frequencies of multiple molecular transitions in distant galaxies and compare against their frequencies today on Earth, \cite{08MuFlMu.mp2me, 11JaXuKl.mp2me} taking advantage of the large time interval\cite{08MuFlMu.mp2me, 16UbBaSa.mp2me} to increase the accuracy of this measurement. A high accuracy constraint on variation in $\mu$ is obtained if the frequency is very accurately known and/or the transition frequency is very sensitive to $\mu$. \Cref{sct:rev} discusses previous constraints of variation in $\mu$ cosmologically using both high abundance molecules and high sensitivity transitions. Measurements of $\mu$ variation on Earth are also being pursued,\cite{08ShBuCh.mp2me,19KoOgIn.mpme} but these constraints are much less tight than astronomical observations (on order $10^{-14}$/yr for Earth-based experiments vs $10^{-17}$/yr for astronomical observations) due to the much shorter time intervals involved.


Thompson was the first to demonstrate how molecular energy levels and transition frequencies are dependent on the proton-to-electron mass ratio, $\mu$. \cite{75Th.mp2me} A shift in $\mu$ will cause a shift in the energy levels in a molecule, propagating through to a shift in observed transition frequencies.  
Each transition has a different sensitivity to $\mu$, quantified by the sensitivity coefficient $K$ of the transition given by
\begin{equation}
    \frac{\Delta \nu}{\nu} = \frac{\nu_{obs}- \nu_{ref}}{\nu_{ref}} = K \frac{\Delta \mu}{\mu},
    \label{eqn:delta_nu}
\end{equation}
where $\frac{\Delta \mu}{\mu}$ is the fractional change in $\mu$, $\frac{\Delta \nu}{\nu}$ is the fractional change in the transition frequencies ($\nu$), $\nu_{obs}$ is the observed transition frequency, and $\nu_{ref}$ is a high accuracy reference transition frequency. Equation \ref{eqn:delta_nu} makes the assumptions that a variation in $\alpha$ and other fundamental constants will have no effect; these assumptions are standard for non-relativistic molecular transitions and is suitable for the purposes of this paper. However, sensitivity to $\alpha$ should be explicitly tested\cite{11BeBoFl.mp2me} for promising transitions. 
As technology evolves, we can measure spectral lines and transition frequencies with increasing accuracy.

Sensitivity coefficients are a theoretical quantity and need to be calculated in order to direct and interpret astrophysical studies. 

There are general trends in the sensitivity coefficient depending on the type of spectral transition. Simplified models yield the sensitivities as $K_{el} = 0$ for pure electronic transitions, $K_{vib} = -\frac{1}{2}$ for pure vibrational transitions, $K_{rot} = -1$ for pure rotational transitions. \cite{14JaBeUb.mp2me} 

Transitions that are a combination of different types of transitions have previously been shown to deviate from these clear cut sensitivities. \cite{13KoLe.mp2me} High sensitivity coefficients have been observed particularly for transitions between accidentally near degenerate energy levels. 

Throughout this paper, we will use the term enhanced to describe any transition with a sensitivity coefficient of $|K|>5$.

For specific molecular transitions, there are two main ways that the sensitivity of a transition is determined in the literature, model Hamiltonians \cite{11LeKoRe.mp2me, 13Ko.mp2me, 09Ko.mp2me, 12IlJaKo.mp2me} and variational programs to calculate the spectra. \cite{15OwYuPo.mp2me, 18OwYuSp.mp2me}
We will utilise the variational method, which has a key advantage of being able to quickly determine the sensitivities for millions of transitions in the same manner. \cite{15OwYuPo.mp2me}  The sensitivity of a transition between lower state, $i$, and upper state, $j$, is usually determined using 
\begin{equation}
    K_{\mu}(i \rightarrow j) = \frac{\mu (\pdv{E_j}{\mu}) - \mu (\pdv{E_i}{\mu})}{E_j - E_i},
    \label{eqn:K}
\end{equation}
where $E$ is the energy of the lower ($E_i$) and upper ($E_j$) levels of the transitions, and $\pdv{E_j}{\mu}$ is the derivative of the energy with respect to $\mu$.  \cite{11JaXuKl.mp2me}

Previous experimental investigations utilising these theoretical $K$ values, detailed in \Cref{sct:rev}, have constrained the proton-to-electron mass variation to be approximately less than $10^{-17}$ per year. These investigations have focused on using molecules with either high abundance like \ce{H2} or with enhanced transitions like \ce{NH3} and \ce{CH3OH}.  To improve upon the existing constraints and subsequently provide more guidance to new theories of physics, new molecular probes may be useful. 

The goal of this field of research is to measure the frequency of a small number of spectral transitions in molecules in distant galaxies and compare their frequency to Earth-based references. Thus, the quality of a measurement of the variation in the proton-to-electron mass ratio, $\mu$, depends on
\begin{itemize}
\item the accuracy with which we can determine the frequency of each transition, 
\item the difference in the sensitivity coefficients, $K$, between the set of transitions measured. 
\end{itemize}

The accuracy with which we can determine the frequency of each transition is determined by
\begin{itemize}
    \item the strength of the transition itself,
    \item the abundance of the molecule in a distant environment and the distance of that environment from Earth, 
    \item the sensitivity and spectral resolution of telescopes available in the relevant spectral range on Earth.
\end{itemize}

Recent proposals for new molecular probes of variation in $\mu$ have generally focused on identifying new molecules with similar symmetry properties to ammonia or methanol that may have very high sensitivity coefficients. These studies have successfully identified enhanced transitions in molecules, usually arising from close-lying or nearly degenerate energy levels. However, the accuracy with which we can determine the frequency of each transition astrophysically is often a secondary concern. 

Here, we take a different approach by first starting with the set of known astrophysical molecules (focusing on extragalactic molecules), then considering this set of molecules in light of the other factors influencing the quality of the desired measurement. \Cref{sct:crit} details the relevant astrophysical observation factors, while \Cref{sct:SM} uses detailed spectroscopic models of rovibronic transitions in 11 diatomic molecules to investigate the key molecular properties that can lead to enhanced transitions. Finally, \Cref{sct:predict} brings together all factors to screen all extragalactically known diatomic molecules as cosmological probes of $\mu$ variation, and provide a brief discussion on the implications of these results to the screening of polyatomic molecules as probes of cosmological $\mu$ variation.

\section{Previous Investigations}\label{sct:rev}


\begin{figure*}
\centering
  \includegraphics[width = 0.7\textwidth]{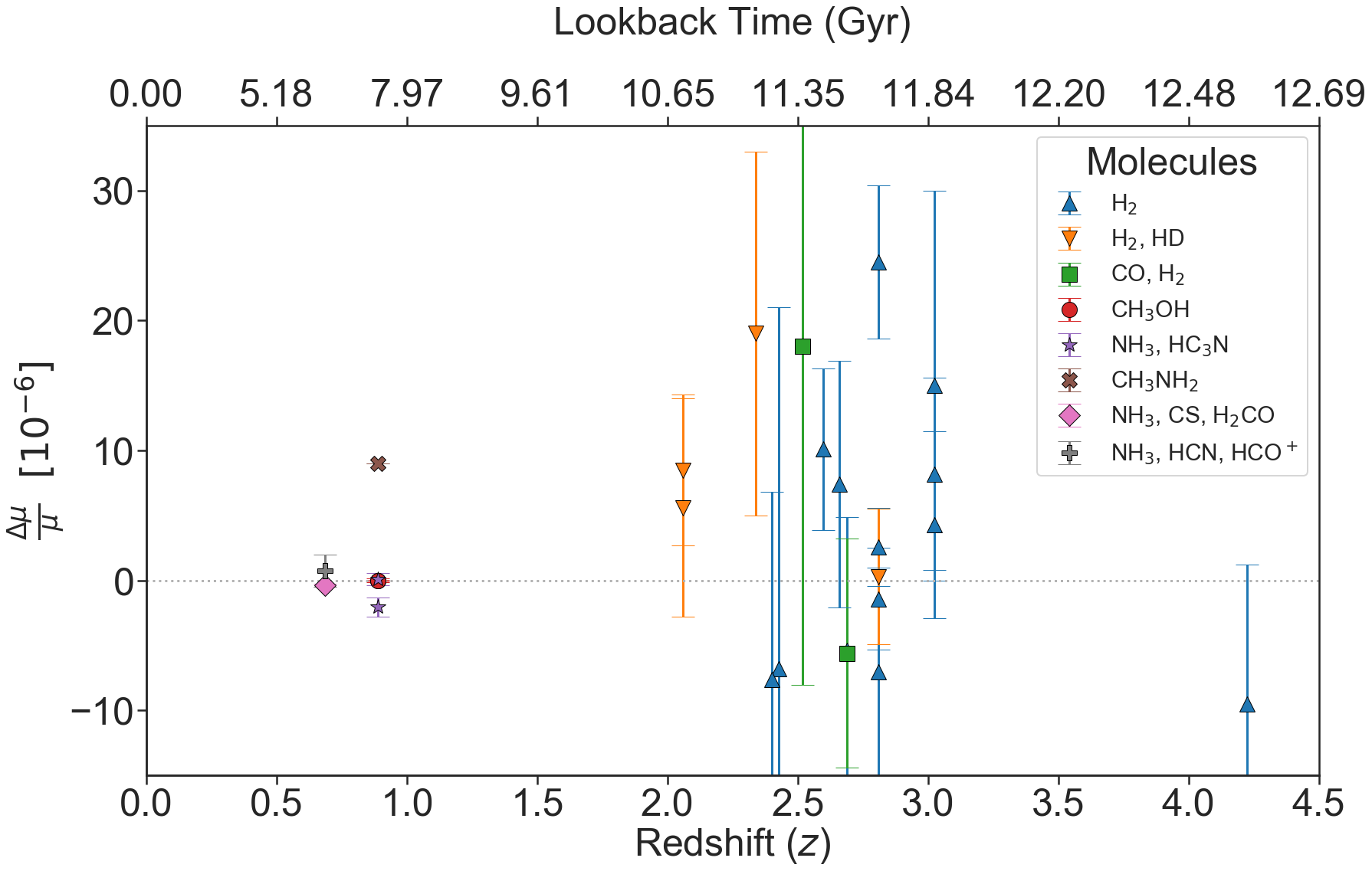}
  \caption{Previous constraints on $\frac{\Delta\mu}{\mu}$ across redshift and look-back time, (using standard LCDM cosmology \cite{12KaLaSt.mp2me}). References for data points of; \ce{H2} constraints \cite{15BaUbMu.mp2me, 11WeMo.mp2me, 12WeMo.mp2me, 08KiWeMu.mp2me, 08WeRe.mp2me, 15DaBaMu.mp2me, 14VaRaNo.mp2me, 08KiWeMu.mp2me, 09ThBeBl.mp2me, 07UbBuEi.mp2me, 08KiWeMu.mp2me, 12BaMuKa.mp2me, 13RaWeSr.mp2me}, \ce{H2} and \ce{HD} constraints \cite{11KiMuUb.mp2me, 17DaVaMu.mp2me, 10MaBuMu.mp2me, 11VaMuMa.mp2me}, \ce{CO} and \ce{H2} constraints \cite{16DaNiSa.mp2me, 17DaNoVo.mp2me} , \ce{NH3} constraints \cite{09HeMeMu.mp2me, 11MuBeGu.mp2me, 11Ka.mp2me, 08MuFlMu.mp2me}, \ce{CH3OH} constraints \cite{15KaUbMe.mp2me,13BaJaHe.mp2me}, and \ce{CH3NH2} constraint \cite{12IlJaKo.mp2me}. It should be noted that the \ce{CH3NH2} constraint is only an upper limit.}
  \label{fig:contraints}
\end{figure*}

\Cref{fig:contraints} shows the constraints that previous experiments have placed on the variation in proton-to-electron mass ratio, $\mu$, over cosmological time. 
These results can be clearly divided into two classes: 
\begin{enumerate} 
\item low sensitivity electronic transitions in common molecules (\ce{H2} and \ce{CO}) observed in very distant galaxies, with a cosmological redshift $z>2$, using optical spectroscopy,
\item high sensitivity transitions observed in less abundant molecules in closer galaxies, with a cosmological redshift $z<1$, using radio spectroscopy.  
\end{enumerate}


\subsection*{High redshift galaxies, low sensitivity}
Molecules with high astrophysical abundance, particularly \ce{H2} and \ce{CO}, are observed in high redshift ($z>2$) galaxies if there are allowed transitions. Initial investigations into possible cosmological $\mu$ variation focused on these systems, taking advantage of the large look-back times in excess of 10 billion years.


\subsubsection*{Molecular Hydrogen \ce{H2}}
Molecular hydrogen is ubiquitous throughout the universe and the most common utilised candidate to constrain a temporal variation of $\mu$, see \Cref{fig:contraints}. \cite{14VaRaNo.mp2me, 12BaMuKa.mp2me, 06MeStIv.mp2me, 07UbBuEi.mp2me, 05IvPeVa.mp2me,16UbKoEi.mp2me, 75Th.mp2me, 16UbBaSa.mp2me, 13BaDaJa.mp2me, 09MaBuMu.mp2me, 11VaMuMa.mp2me, 08KiWeMu.mp2me} With the vibrational and rotational bands forbidden, the observed transitions are strong electric dipole allowed Lyman and Werner absorption transitions between electronic states that are naturally in the UV spectral region, but are shifted into the range of ground-based observations above wavelength of 3050 \AA at large redshift, $z>2$. \cite{11Ch.mp2me} 
As electronic transitions, the sensitivity coefficients of these bands are extremely low, with $|K| < 0.05$. \cite{06MeStIv.mp2me} 
Despite this, the high abundance of \ce{H2} means that it is still useful for constraining a variation of $\mu$. \cite{16UbBaSa.mp2me} A weighted average of \ce{H2} results shown in \Cref{fig:contraints} from redshift $z=2.0-4.2$ gives $\Delta\mu/\mu = (3.1 \pm 1.6) \times 10^{-6}$. \cite{18Ub.mp2me}

\subsubsection*{Carbon Monoxide \ce{CO}}
\ce{CO} is the second most abundant molecule in gas form in the universe, and has been studied extensively with laboratory spectroscopy.  \cite{16DaNiSa.mp2me} Following a proposal in 2012,\cite{12SaNiBa.mp2me} transitions from the electronic A-X system has been used to supplement \ce{H2} measurements. The sensitivities of these transitions are low with $- 0.002 < K < 0.066$ \cite{12SaNiBa.mp2me} and provide less stringent constraints than that determined by \ce{H2} systems. \cite{16DaNiSa.mp2me} A combined result using both \ce{CO} and \ce{H2} from $z_{abs} = 2.69$ towards background quasar SDSS J1237+0647 gives $\Delta\mu/\mu = (-5.6 \pm 5.6_{stat} \pm 3.1_{syst}) \times 10^{-6}$ which is consistent with a null result. \cite{16DaNiSa.mp2me} 

The vibrational and rotational bands of CO have not been considered for placing a cosmological constraint on the variation of the proton-to-electron mass ratio. These lower frequency bands should have higher sensitivity coefficients and may be an interesting avenue of further exploration if appropriate telescopes are available in the red-shifted spectral range of these transitions.  

\subsection*{Low redshift galaxies, enhanced sensitivity}
Recently, radio observations in low redshift $z<1$ galaxies have allowed for much tighter constraints to be placed on a variation of $\mu$. We can once again see this in \Cref{fig:contraints} with constraints placed by observations of molecular lines observed from galaxies with look-back times of less than 8 billion years. 

\subsubsection*{Ammonia \ce{NH3}}
Ammonia, \ce{NH3}, is one of the most abundant polyatomic molecules observed in the interstellar medium, however, only 2 extragalactic sources have been found and both are in low redshift galaxies. \cite{13KoLe.mp2me} 
The inversion transition near 24 GHz has a sensitivity coefficient of $K = 4.46$ and is generally compared to the rotational transitions of other molecules which are present at a similar frequency. \cite{08MuFlMu.mp2me} 
This approach gives very high accuracy constraints.  However, the use of multiple molecules in the comparison introduces the potential for systematic error as the assumption must be made that the gas of these different molecules co-exist in the same molecular gas cloud and have the same velocity structure. \cite{13BaJaHe.mp2me} 


There have been suggestions of using mixed transitions in the isotopologues of ammonia. 
\ce{NH2D} and \ce{NHD2} have been considered for their mixed rotation-inversion transitions, as they can limit systematic errors. However, the observed spectral transitions of these isotopologues have small sensitivities. \cite{13Ko.mp2me}

\subsubsection*{Methanol}


Mixed rotation-torsional transitions in methanol are significantly more sensitive than ammonia or hydrogen. Using multiple rotation-torsional methanol transitions allows for a $\Delta K \approx 60$, \cite{11LeKoRe.mp2me} which is an order of magnitude greater than achievable with ammonia, and 1000 times better than that of \ce{H2}. \cite{11LeKoRe.mp2me} 



Methanol has been commonly observed within our galaxy, but only one extragalactic system has observed methanol; PKS1830-211 at redshift $z = 0.89$. \cite{11MuBeGu.mp2me} There has been some discussion about the observed system with methanol, and if all the observed transitions  do indeed come from the same gas cloud. \cite{15KaUbMe.mp2me} This has encouraged more rigorous analysis to ensure spectral lines are coming from the same environment. With the transitions that have been observed the tightest temporal constraint has been placed on $\frac{\Delta\mu}{\mu}$ of $(-2.9 \pm 5.7) \times 10^{-8}$, which is a variation of $\dv{\mu}{t}/\mu < 2 \times 10^{-17} \text{yr}^{-1}$ assuming a linear change in time. \cite{15KaUbMe.mp2me}


    

\subsubsection*{Other proposed molecules}
Aside from the molecules used in placing constraints, many different molecules have also been proposed as possible probes for testing a variation in $\mu$.  \Cref{tbl:proposed} shows many of the molecules proposed as astrophysical probes, along with the relevant references. A significant number of the diatomics and some polyatomics in this table have been proposed due to the presence of $\Pi$ states and lambda-doubling \cite{09Ko.mp2me, 13KoLe.mp2me}; this is discussed in further detail in \Cref{sct:lambda}. Symmetry similar to that of ammonia and methanol have also inspired attention to many of the polyatomic suggestions. \cite{11JaKlXu.mp2me, 18OwYuSp.mp2me}

\begin{table*}
\centering
\caption{Table of sensitivity ranges for transitions in molecules that have been used or proposed as probes to astrophysically constrain the variation of the proton-to-electron mass ratio. Those in bold have been used to place constraints.}\label{tbl:proposed}
\begin{tabular}{@{}lllll@{}}
\toprule
\textbf{Molecules}            & Electronic State                                                                & Type of transition         & Range of sensitivity  & Reference                             \\ \midrule
\textbf{Diatomic molecules}   &                                                                                 &                            &                       &                                       \\ \cmidrule(r){1-1}
\textbf{\ce{H2}}                       & $B^1 \Sigma^+_u \leftarrow X^1 \Sigma^+_g /C^1 \Pi_u \leftarrow X^1 \Sigma^+_g$ & $E_{el}/E_{vib}$           & $-0.054 < K < +0.019$ & \cite{07UbBuEi.mp2me, 06MeStIv.mp2me} \\
\textbf{\ce{HD}}                       & $B^1 \Sigma^+_u \leftarrow X^1 \Sigma^+_g /C^1 \Pi_u \leftarrow X^1 \Sigma^+_g$ & $E_{el}/E_{vib}$           & $-0.052 < K < +0.012$ & \cite{08IvRoVi.mp2me}                 \\
\textbf{\ce{CO} }                      & $A^1 \Pi \leftarrow X^1\Sigma^+$                                                & $E_{el}/E_{vib}$           & $-0.071 < K < +0.003$ & \cite{12SaNiBa.mp2me}                 \\
\ce{CH}                       & $X^2 \Pi$                                                                       & $E_{fs}/E_{rot}$        & $-6.2 < K < +2.7$     & \cite{09Ko.mp2me, 12DeUbBe.mp2me}     \\
\ce{CD}                       & $X^2 \Pi$                                                                       & $E_{fs}/E_{rot}$        & $-67 < K < +18$       & \cite{12DeUbBe.mp2me}                 \\
\ce{OH}                       & $X^2 \Pi$                                                                       & $E_{fs}/E_{rot}$        & $-460 < K < -0.50$    & \cite{09Ko.mp2me}                     \\
\ce{NO}                       & $X^2 \Pi$                                                                       & $E_{fs}/E_{rot}$        & $-38.9 < K < +6.81$   & \cite{09Ko.mp2me}                     \\
\ce{LiO}                      & $X^2 \Pi$                                                                       & $E_{fs}/E_{rot}$        & $-4.24 < K < -0.95$   & \cite{09Ko.mp2me}                     \\
\ce{NH+}                      & $a4\Sigma^{-} \leftarrow X^2 \Pi (v = 0, 1)$                                    & $E_{el}/E_{rot}$           & $-185.8 < K < +126.9$ & \cite{11BeKoBo.mp2me}                 \\
                              &                                                                                 &                            &                       &                                       \\
\textbf{Polyatomic molecules} &                                                                                 &                            &                       &                                       \\ \cmidrule(r){1-1}
\ce{\textit{l}-C3H}           & $X^2 \Pi$                                                                       & $E_{vib}/E_{rot}$      & $-19 < K < +742$      & \cite{13Ko.mp2me}                     \\
\textbf{\ce{NH3}}                      & $X$                                                                    & $E_{inv}$                  & $-4.2$                & \cite{07FlKo.mp2me, 15OwYuTh.mp2me}                   \\
\ce{ND3}                      & $X$                                                                             & $E_{inv}$                  & $-5.6$                & \cite{04VeKuBe.mp2me, 15OwYuTh.mp2me}                 \\
                            &     $X$                                                 &	$E_{rot}/E_{vib}$           &	$-32 < K < +28$  &	\cite{16OwYuTh.mp2me} \\
\ce{NH2D} and \ce{ND2H}       & $X$                                                                             & $E_{inv}/E_{rot}$          & $-1.54 < K < +0.10$   & \cite{10KoLaLe.mp2me}                 \\
\ce{H3O+}                     & $X$                                                                             & $E_{inv}$                  & $-2.5$                & \cite{11KoLe.mp2me, 16OwYuTh.mp2me}   \\
                              & $X$                                                                             & $E_{inv}/E_{rot}$          & $-9.0 < K < +5.7$     & \cite{11KoLe.mp2me, 15OwYuPo.mp2me}                   \\
\ce{H2DO+} and \ce{D2HO+}     & $X$                                                                             & $E_{inv}/E_{rot}$          & $-219 < K < +11.0$    & \cite{11KoPoRe.mp2me}                 \\
\ce{D3O+}	                   &    $X$	                                                                         & $E_{inv}/E_{rot}$	      &$-7.5 < K < + 3.5$    &	\cite{11KoPoRe.mp2me, 15OwYuPo.mp2me} \\
\ce{H2O2}                     & $X$                                                                             & $E_{inv}/E_{rot}$          & $-36.5 < K < +13.0$   & \cite{11Ko.mp2me}                     \\
\ce{PH3}                      & $X$                                                                             & $E_{rot}/E_{vib}$          & $-32 < K < +461$      & \cite{18OwYuSp.mp2me}                 \\
\textbf{\ce{CH3OH} }                   & $X$                                                                             & $E_{tors}/E_{rot}$         & $-88 < K < +330$      & \cite{11JaXuKl.mp2me, 11LeKoRe.mp2me} \\
\ce{CH3SH}                    & $X$                                                                             & $E_{tors}/E_{rot}$         & $-14.8 < K < +12.2$   & \cite{13JaXuKl.mp2me}                 \\
\ce{CH3COH}                   & $X$                                                                             & $E_{tors}/E_{rot}$         & $-3.7 < K < -0.5$     & \cite{11JaKlXu.mp2me}                 \\
\ce{CH3CONH2}                 & $X$                                                                             & $E_{tors}/E_{rot}$         & $-1.34 < K < +0.06$   & \cite{11JaKlXu.mp2me}                 \\
\ce{HCOOCH3}                  & $X$                                                                             & $E_{tors}/E_{rot}$         & $-1.07 < K < -0.03$   & \cite{11JaKlXu.mp2me}                 \\
\ce{CH3COOH}                  & $X$                                                                             & $E_{tors}/E_{rot}$         & $-1.36 < K < -0.27$   & \cite{11JaKlXu.mp2me}                 \\
\textbf{\ce{CH3NH2} }                  & $X$                                                                             & $E_{inv}/E_{tors}/E_{rot}$ & $-19 < K < +24$       & \cite{12IlJaKo.mp2me}                 \\ 
\ce{(CH3)2CO}	            &  	$X$                             & $E_{rot}/E_{tors}$        &	$-1.98 < K < +6.07$      & 	\cite{14Il.mp2me}\\
\ce{C2H6O2}	            &     $X$                 &  	$E_{tun}/E_{rot}$	            &   $-17.9 < K < +16.5$	        &\cite{14ViKo.mp2me}\\
\bottomrule
\end{tabular}
\end{table*}


\begin{table*}
\centering
\caption{All diatomic molecules observed in the interstellar medium and circumstellar shells \cite{18Mc.mp2me} with ground state and first excited electronic state information, approximate abundance relative to \ce{H2} or relative to atomic H if denoted with **. LL - published line-list available through the ExoMol database (http://exomol.com/), SM - published spectroscopic model for nuclear motion program. All electronic state information taken from \cite{79HeHu.diatomic} except for \ce{ScH} taken from \cite{15LoYuTe.exomol}, and \ce{VO} taken from \cite{16McYuTe.exomol}.}\label{tbl:diatomics}
\resizebox{\textwidth}{!}{%
\begin{tabular}{lccrcrclll}
\toprule
\textbf{Diatomic} & \textbf{Ground State} & \textbf{\begin{tabular}[c]{@{}l@{}}First Excited \\ state\end{tabular}} & \textbf{\begin{tabular}[c]{@{}l@{}}First \\ $T_e$ (\cm{})\end{tabular}} & \textbf{\begin{tabular}[c]{@{}l@{}}First Allowed \\ Excited State\end{tabular}} & \textbf{\begin{tabular}[c]{@{}l@{}}Allowed \\ $T_e$ (\cm{})\end{tabular}} & \textbf{LL} & \textbf{SM} & \textbf{\begin{tabular}[c]{@{}l@{}}Relative\\ Abundance\\ to \ce{H2} or H**\end{tabular}} & \textbf{Ref} \\
\midrule
\multicolumn{2}{l}{\textbf{Interstellar and circumstellar}}  &&&&&&&&                                  \\ \cmidrule(r){1-2}
\ce{AlCl} & $X^1\Sigma^+$ & $a^3\Pi$ & 24528 & $A^1\Pi$ & 38254 & Y &  & $10^{-8}$ & \cite{10TeZi.abun} \\
\ce{AlF} & $X^1\Sigma^+$ & $a^3\Pi$ & 27241 & $A^1\Pi$ & 43949 & Y &  &  &  \\
\ce{AlO} & $X^2\Sigma^+$ & $A^2\Pi$ & 5341 & $A^2\Pi$ & 5341 & Y & \textsc{Duo} & $10^{-8}$ & \cite{10TeZi.abun} \\
\textbf{\ce{ArH+}} & $X^1\Sigma^+$ &  &  &  &  & N &  & $10^{-10}$** & \cite{16GeNeGo.abun} \\
\ce{C2} & $X^1\Sigma^+$ & $a^3\Pi$ & 716 & $A^1\Pi$ & 8391 & Y & \textsc{Duo} & $10^{-8}$ & \cite{13McWaMa.abun, 13Ti.abun} \\
\textbf{\ce{CF+}} &  &  &  &  &  & N &  & $10^{-10}$ & \cite{16MuKaBl.abun} \\
\textbf{\ce{CH}} & $X^2\Pi$ & $a^4\Sigma^-$ & 5844 & $A^2\Delta$ & 23189 & Y &  & $10^{-8}$ & \cite{13McWaMa.abun, 13Ti.abun, 16GeNeGo.abun} \\
\textbf{\ce{CH+}} & $X^1\Sigma^+$ & $a^3\Pi$ & 9200 & $A^1\Pi$ & 24111 & N &  & $10^{-8}$ & \cite{13Ti.abun, 16GeNeGo.abun} \\
\textbf{\ce{CN}} & $X^2\Sigma^+$ & $A^2\Pi$ & 9245 & $A^2\Pi$ & 9245 & Y &  & $10^{-9}$ & \cite{13McWaMa.abun, 13Ti.abun} \\
CN$^-$ &  &  &  &  &  & N &  & $10^{-9}$ & \cite{10AgCeGu.abun} \\
\textbf{\ce{CO}} & $X^1\Sigma^+$ & $a^3\Pi$ & 48686 & $A^1\Pi$ & 65075 & Y &  & $10^{-5}$ & \cite{13McWaMa.abun,13Ti.abun} \\
\textbf{\ce{CO+}} & $X^2\Sigma^+$ & $A^2\Pi$ & 20733 & $A^2\Pi$ & 20733 & N &  & $10^{-15}$ & \cite{98CeCaWo.abun} \\
\ce{CP} & $X^2\Sigma^+$ & $A^2\Pi$ & 6895 & $A^2\Pi$ & 6895 & Y &  & $10^{-8}$ & \cite{08MiHaTe.abun} \\
\textbf{\ce{CS}} & $X^1\Sigma^+$ & $a^3\Pi$ & 27661 & $A^1\Pi$ & 38904 & Y & LEVEL & $10^{-9}$ & \cite{13McWaMa.abun, 13Ti.abun} \\
\ce{FeO} & $X^5\Delta$ & $A^5\Sigma^+$ & 3948 & $B^5\Pi$ & 14404 & N &  & $10^{-11}$ & \cite{03FuWaNa.abun} \\
\ce{H2} & $X^1\Sigma^+$ & $B^1\Sigma^+$ & 91700 & $B^1\Sigma^+$ & 91700 & N &  & $1$ & \cite{13Ti.abun} \\
\ce{HCl} & $X^1\Sigma^+$ & $A^1\Pi$? & 76322 &  &  & Y &  & $10^{-10}$ & \cite{13Ti.abun, 16GeNeGo.abun} \\
\ce{HCl+} & $X^2\Pi$ & $A^2\Sigma^+$ & 28626 & $A^2\Sigma^+$ & 28626 & N &  & $10^{-9}$** & \cite{16GeNeGo.abun} \\
\textbf{\ce{HD}} & $X^1\Sigma^+$ & $B^1\Sigma^+$ & 91698 & $B^1\Sigma^+$ & 91698 & N &  & $10^{-7}$ & \cite{13Ti.abun} \\
\textbf{\ce{HF}} & $X^1\Sigma^+$ & $B^1\Sigma^+$ & 84776 & $B^1\Sigma^+$ & 84776 & Y &  & $10^{-8}$ & \cite{16GeNeGo.abun} \\
\ce{KCl} & $X^1\Sigma^+$ &  &  &  &  & Y & LEVEL &  &  \\
\ce{N2} & $X^1\Sigma^+$ & $A^3\Sigma^+$ & 50203 & $a^1\Sigma^-$ & 68152 & N &  & $10^{-7}$ & \cite{13Ti.mp2me} \\
\ce{NaCl} & $X^1\Sigma^+$ &  &  &  &  & Y & LEVEL & $10^{-7}$ & \cite{17PrSaCe.abun} \\
\textbf{\ce{NH}} & $X^3\Sigma^-$ & $a^1\Delta$ & 12566 & $a^3\Pi$ & 29807 & Y &  & $10^{-9}$ & \cite{13Ti.abun, 16GeNeGo.abun} \\
\textbf{\ce{NO}} & $X^2\Pi$ & $a^4\Pi$ & 38440 & $A^2\Sigma^+$ & 43965 & Y & \textsc{Duo} & $10^{-8}$ & \cite{13McWaMa.abun} \\
\ce{NO+} & $X^1\Sigma^+$ & $A^3\Sigma^+$ & 52190 & $A^1\Pi$ & 73471 & N &  & $10^{-10}$ & \cite{14CeBaAl.abun} \\
\textbf{\ce{NS}} & $X^2\Pi$ & $B^2\Pi$ & 30294 & $B^2\Pi$ & 30294 & Y & \textsc{Duo} & $10^{-9}$ & \cite{03MaMaMa.abun} \\
\ce{NS+} & $X^1\Sigma^+$ &  &  &  &  & N &  & $10^{-10}$ & \cite{18CeLeAg.abun} \\
\ce{O2} & $X^3\Sigma^-$ & $a^1\Delta$ & 7918 & $A^3\Sigma^+$ & 35397 & N &  & $10^{-8}$ & \cite{13McWaMa.abun} \\
\textbf{\ce{OH}} & $X^2\Pi$ & $A^2\Sigma^+$ & 32684 & $A^2\Sigma^+$ & 32684 & Y &  & $10^{-7}$ & \cite{13McWaMa.abun,13Ti.abun, 16GeNeGo.abun} \\
\textbf{\ce{OH+}} & $X^3\Sigma^-$ & $a^1\Delta$ & 17660 & $a^3\Pi$ & 28438 & Y &  & $10^{-8}$ & \cite{13Ti.abun} \\
\ce{PN} & $X^1\Sigma^+$ & $A^1\Pi$ & 39805 & $A^1\Pi$ & 39805 & Y & LEVEL & $10^{-9}$ & \cite{08MiHaTe.abun} \\
\ce{PO} & $X^2\Pi$ & $B^2\Sigma^+$ & 30730 & $B^2\Sigma^+$ & 30730 & Y & \textsc{Duo} & $10^{-6}$ & \cite{17PrSaCe.abun} \\
\ce{SH} & $X^2\Pi$ & $A^2\Sigma^+$ & 31038 & $A^2\Sigma^+$ & 31038 & Y & \textsc{Duo} & $10^{-8}$ & \cite{16GeNeGo.abun} \\
\textbf{\ce{SH+}} & $X^3\Sigma^-$ & $a^3\Pi$ &  &  &  & N &  & $10^{-8}$** & \cite{16GeNeGo.abun} \\
\ce{SiC} & $X^3\Pi$ & $A^3\Sigma^-$ & 5597 & $A^3\Sigma^-$ & 5597 & N &  &  &  \\
\ce{SiH} & $X^2\Pi$ & $A^2\Delta$ & 24300 & $A^2\Delta$ & 24300 & Y & \textsc{Duo} &  &  \\
\ce{SiN} & $X^2\Sigma^+$ & $A^2\Pi$ & 8000 & $A^2\Pi$ & 8000 & N &  & $10^{-7}$ & \cite{03ScLeMe.abun} \\
\textbf{\ce{SiO}} & $X^1\Sigma^+$ & $A^3\Sigma^+$ & 33630 & $C^1\Sigma^-$ & 38624 & Y & LEVEL & $10^{-6}$ & \cite{17PrSaCe.abun} \\
\ce{SiS} & $X^1\Sigma^+$ & $a^3\Pi$ & 30239 & $D^1\Pi$ & 35026 & Y & LEVEL & $10^{-6}$ & \cite{17PrSaCe.abun} \\
\textbf{\ce{SO}} & $X^3\Sigma^-$ & $a^1\Delta$ & 6350 & $a^3\Pi$ & 38306 & N &  & $10^{-8}$ & \cite{13McWaMa.abun, 13Ti.abun} \\
\textbf{\ce{SO+}} & $X^2\Pi$ & $a^4\Pi$ & 26170 & $A^2\Pi$ & 35600 & N &  & $10^{-10}$ & \cite{11MuBeGu.mp2me} \\
\ce{TiO} & $X^3\Delta$ & $a^1\Delta$ & 3440 & $E^3\Pi$ & 12025 & Y & \textsc{Duo} &  & \\
\vspace{0.5em} \textbf{Other}  &&&&&&&&&                                  \\ \cmidrule(r){1-1}
\ce{PS}	&$X^2\Pi$&	$B^2\Pi$&	22894&	$B^2\Pi$&	22894&	Y &	\textsc{Duo} & & \\
\ce{ScH} &	$X^1\Sigma^+$ &	$b^3\Pi$ &	4152 &	$B^1\Pi$ &	5856 &	Y &	\textsc{Duo} & & \\
\ce{VO}	 & $X^4\Sigma^-$ &	$A^4\Phi$ &	7293 & 	$a^4\Pi$ &	9561 &	Y &	\textsc{Duo} & & \\
\bottomrule
\end{tabular}%
}
\end{table*}

\begin{table}
\centering
\caption{Small (N<8) polyatomic molecules that have been observed in extragalactic environments. \cite{18Mc.mp2me} \ce{H2CS} is a tentative detection. Those in bold have line lists available as a part of the ExoMol database. }\label{tbl:poly}
\resizebox{0.45\textwidth}{!}{%
\begin{tabular}{@{}lllll@{}}
\toprule
\textbf{3 atoms} & \textbf{4 atoms} & \textbf{5 atoms} & \textbf{6 atoms} & \textbf{7 atoms} \\ \midrule
\textbf{\ce{H2O}} & \textbf{\ce{H2CO}} & \ce{c-C3H2} & \ce{CH3OH} & \ce{CH3CCH} \\
\textbf{\ce{HCN}} & \textbf{\ce{NH3}} & \ce{HC3N} & \ce{CH3CN} & \ce{CH3NH2} \\
\ce{HCO+} & \ce{HNCO} & \ce{CH2NH} & \ce{HC4H} & \ce{CH3CHO} \\
\ce{C2H} & \ce{C2H2} & \ce{NH2CN} & \ce{HC(O)NH2} &  \\
\ce{HNC} & \ce{H2CS} & \ce{l-C3H2} &  &  \\
\ce{N2H+} & \ce{HOCO+} & \ce{H2CCN} &  &  \\
\ce{OCS} & \ce{c-C3H} & \ce{H2CCO} &  &  \\
\ce{HCO} & \ce{H3O+} & \ce{C4H} &  &  \\
\textbf{\ce{H2S}} & \ce{l-C3H} &  &  &  \\
\textbf{\ce{SO2}} &  &  &  &  \\
\ce{HOC+} &  &  &  &  \\
\ce{C2S} &  &  &  &  \\
\ce{H2O+} &  &  &  &  \\
\ce{HCS+} &  &  &  &  \\
\ce{H2Cl+} &  &  &  &  \\
\ce{NH2} &  &  &  &  \\
\textbf{\ce{H3+}} &  &  &  &  \\
\ce{C3} &  &  &  &  \\ \bottomrule
\end{tabular}%
}
\end{table}

\section{Astrophysical criteria}\label{sct:crit}
The first and most necessary criteria for a potential cosmological probe of the proton-to-electron mass ratio, $\mu$, variation is the existence of the molecule astrophysically, ideally extragalactically.  The subset of  molecules that have been observed in space is tiny compared to all of the molecules on Earth.

\Cref{tbl:diatomics} lists all of the diatomics which have been observed in the interstellar medium, circumstellar shells or extragalactically \cite{18Mc.mp2me}, together with some other information which we will discuss further in \cref{ssct:res} and \cref{sct:predict}. Just 19 diatomic molecules have been observed extragalactically (shown in bold in \Cref{tbl:diatomics}).  Of the other 125 small (N<8) polyatomic molecules which have been detected within the interstellar medium, \cite{18Mc.mp2me} \Cref{tbl:poly} shows the 42 that have been observed extragalactically. 

The abundance of the molecule is also an important factor; more abundant molecules will give stronger spectral signals. \Cref{tbl:diatomics} provides estimates for galactic abundance of many diatomics; however, it is crucial to note that this is highly environment dependent quantity and not well constrained currently. 

The third criteria is that we have a telescope that is available to measure molecular spectral lines (taking into account redshift). For example, though \ce{H2} is the most abundant molecule astrophysically, we are unable to directly observe \ce{H2} at distances closer than redshift $z=2$ (i.e. about 10 billion years ago) because the vibrational and rotational transitions are dipole forbidden and thus extremely weak. 

\begin{figure}
    \centering
    \includegraphics[width=0.5\textwidth]{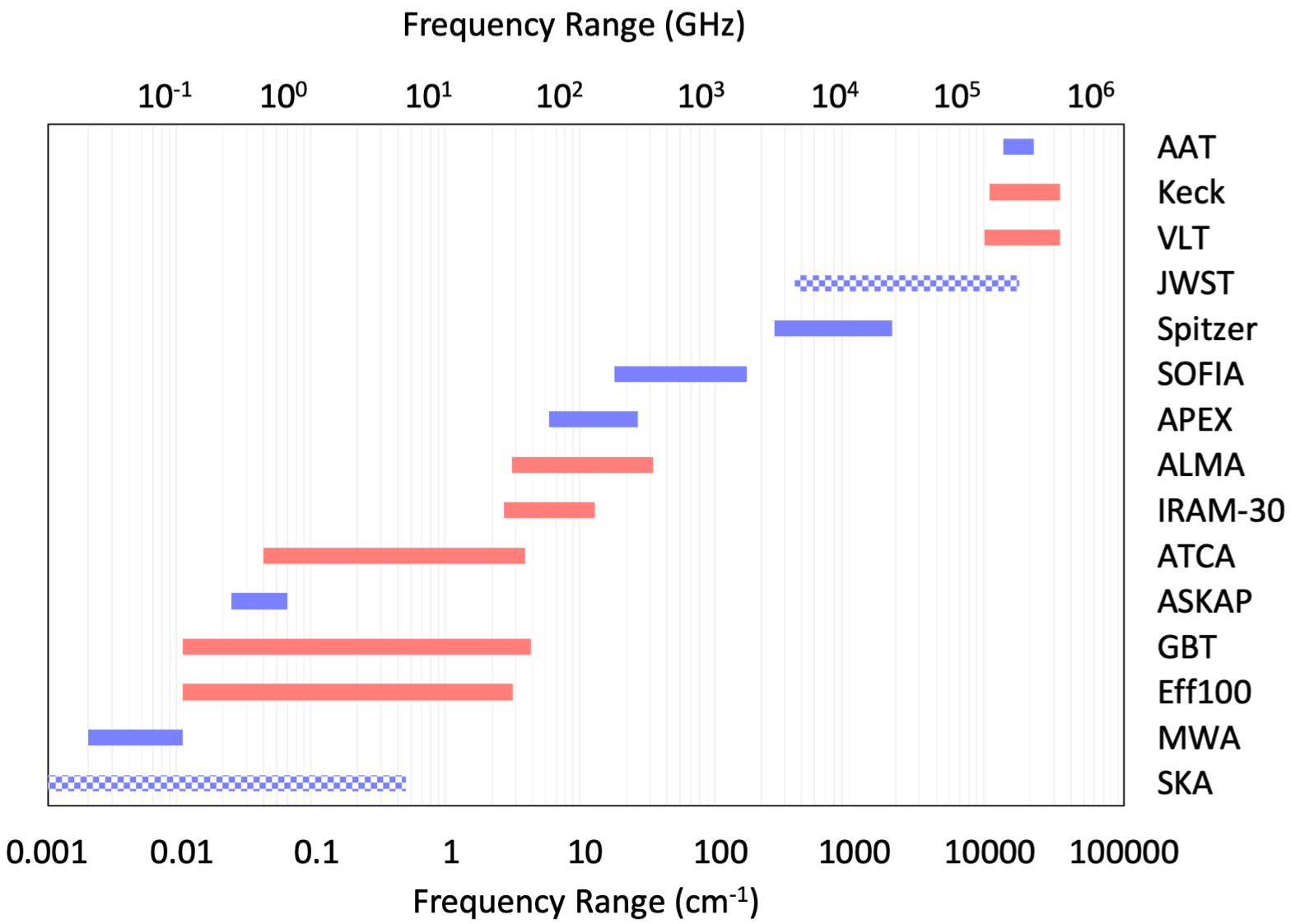}
    \caption{The frequency ranges in both \cm and GHz of some of the telescopes which have been used or could be used to observe molecular lines. The telescopes in red have contributed observations to place a constraint on the variation of the proton-to-electron mass ratio. Common acronyms for the telescopes are used here. Those in blue are potential other telescopes with JWST and SKA yet to be observing in a hatched pattern. }
    \label{fig:telescopes}
\end{figure}

\Cref{fig:telescopes} shows the frequency range of some telescopes that have been used or may be used to constrain a variation in $\mu$. Observing in the optical range and radio range is quite easy from the ground or space, with many available telescopes. The importance of the infrared region to molecular spectroscopy of nearby systems means that there is usually a space-based observation of high quality available. However, historically there have been few telescopes in the far-infrared (between $10^{2} - 10^{4}$ \cm) and in the ultraviolet as this region has few spectral transitions and the telescopes must be space-based. This lack of telescopes in the far-infrared region could be problematic  for some potential probe transitions as it is the unusual transitions between nearly degenerate levels that tend to have very high sensitivity coefficients and these transitions often occur in this region. 

Finally, we must consider the intensity of the transition to be observed, which depends on its intrinsic strength (i.e. its Einstein A coefficient) and the population of the lower energy state (i.e. the temperature of the system). Unfortunately, the conditions that lead to enhanced transitions (mixing of states, close-lying levels) generally occurs higher in the molecular potential energy surface, meaning the enhanced transitions often have low intensities due to Boltzmann factors. Furthermore, molecular gaseous environments astrogalactically are often very cool, on the order of 4-100 K. 

Quantifying the intensity of a transition that can be measured by a telescope is of course always an exercise in approximation; longer observation runs, closer targets and larger telescopes all improve the quality of a measurement. However, within the typical units used by astronomers, transitions with intensities above 10$^{-18}$ cm/molecule may be observable in extragalactic environments, while transitions with intensities less than 10$^{-25}$ and definitely 10$^{-30}$ cm/molecule are unobservable extragalactically for our desired measurements. 

\section{Sensitivity Coefficients in Rovibronic Transitions of Diatomics}\label{sct:SM}

Rovibronic transitions have not been explored extensively in the context of variation in the proton-to-electron mass ratio, $\mu$, cosmologically, with most studies limited to the low sensitivity transitions in \ce{H2}, \ce{HD} and \ce{CO}. The main relevant previous work is the promising sensitivity coefficients found in \ce{NH+} of up to $K=126.9$,\cite{11BeKoBo.mp2me} which is attributed to the presence of an anomolous low-lying electronic excited state in \ce{NH+} only a few hundred \cm{} above the ground state. However, \ce{NH+} has not yet been astrophysically observed. 

To explore the ubiquity of enhanced transitions in rovibronic transitions, here we take advantage of the high quality spectroscopic data available on a set of 11 diatomic molecules, tabulated in \Cref{tbl:SM}. These line lists were initially developed to serve the astrophysical stellar and exoplanet communities.\cite{16TeYuAl.exomol} For each molecule, the available data consists of a molecular line list (set of energy levels and transition frequencies) and, importantly for our purposes, a spectroscopic model that includes the potential energy, dipole moment and coupling curves that produce this line list. It is straightforward to modify the spectroscopic model to mimic the effect of changing $\mu$. We use the \textsc{Duo} nuclear motion package to recalculate the energy levels and thereby extract the sensitivity coefficient of all the molecular transition for a given molecule. 

\begin{table*}
\sisetup{round-mode=places}
\centering
\caption{The 11 different diatomic molecules analysed within this article, including the three different isotopologues of \ce{AlO}. The number of electronic states included in the spectroscopic model, with the total number of transitions, number of enhanced transitions with $|K|>5$, sensitivity of the most sensitive transition, and the maximum difference in sensitivity between two transitions, where the transitions are constrained to intensities of greater than $10^{-30}$ cm/molecule at 1000 K. }\label{tbl:SM}
\begin{tabular}{ccrrS[round-precision=2]S[round-precision=2]c}
\toprule
\textbf{Molecule} & \textbf{\begin{tabular}[c]{@{}l@{}}Electronic \\States\end{tabular}} & \textbf{\begin{tabular}[c]{@{}l@{}}Transitions\end{tabular}} & \textbf{\begin{tabular}[c]{@{}l@{}}Enhanced \\Transitions \end{tabular}} & \textbf{Max(|K|)} & \textbf{Max(|$\Delta$K|)} & \textbf{\begin{tabular}[c]{@{}l@{}}Reference for \\ Spectroscopic Model\end{tabular}} \\ \midrule
\ce{^{32}S^{1}H} & 1 & 19683 & 0 & 2.868946523 & 4.673579174 & \cite{18YuBoGo.exomol} \\
\ce{^{14}N^{16}O} & 1 & 19968 & 0 & 1.011742414 & 1.553239728 & \cite{17WoYuBe.exomol} \\
\ce{^{14}N^{32}S} & 1 & 24484 & 0 & 1.001648498 & 1.150299077 & \cite{18YuBoGo.exomol} \\
\ce{^{31}P^{16}O} & 1 & 28635 & 0 & 1.002250288 & 1.221486032 & \cite{17PrJaLo.exomol} \\
\ce{^{31}P^{32}S} & 3 & 288390 & 0 & 1.002733805 & 1.522273387 & \cite{17PrJaLo.exomol} \\
\ce{^{26}Al^{16}O}&	3 &	597035 &	3974 &	2389.929595	&3773.715055	&\cite{15PaYuTe.exomol} \\
\ce{^{27}Al^{16}O}&	3 &	600540 &	3948 &	3149.960888	&5486.356463	&\cite{15PaYuTe.exomol} \\
\ce{^{27}Al^{18}O}&	3 &	647858 &	4026 &	1105.054115	&1458.562614	&\cite{15PaYuTe.exomol} \\
\ce{^{28}Si^{1}H} & 4 & 139444 & 1 & 5.637553998 & 9.694847249 & \cite{18YuSiLo.exomol} \\
\ce{^{45}Sc^{1}H} & 6 & 318985 & 2300 & 1012.39876 & 1340.061544 & \cite{15LoYuTe.exomol} \\
\ce{^{12}C2} & 8 & 276117 & 984 & 37.39829295 & 46.18420234 & \cite{18YuSzPy.exomol} \\
\ce{^{48}Ti^{16}O} & 11 & 1986433 & 0 & 4.808677391 & 8.598804896 & \cite{19McMaHo.exomol} \\
\ce{^{51}V^{16}O} & 13 & 3447739 & 2302 & 636.8424878 & 1156.225041 & \cite{16McYuTe.exomol} \\ \bottomrule
\end{tabular}
\end{table*}

\subsection{Method}

The advantage of utilising spectroscopic models lies in its ability to trivially change the molecular mass of the system. Assuming all baryonic matter can be treated in the same way we can treat a variation in the molecular mass as a variation in the proton-to-electron mass ratio. \cite{07De.mp2me} 

To calculate the sensitivity coefficients of each transition we are taking a variational approach, similar to that used for investigations of sensitivity in rovibrational transitions of polyatomic molecules. \cite{15OwYuPo.mp2me,15OwYuTh.mp2me,16OwYuTh.mp2me,18OwYuSp.mp2me} Using the variational nuclear motion program \textsc{Duo},\cite{16YuLoTe.mp2me} we can solve the full rovibrationic wavefunction to give energy levels and dipole allowed transitions for diatomics. See ExoMol papers\cite{16YuLoTe.mp2me,17TeYu.exomol} for the method of calculating transitional frequencies and Einstein A coefficients. To calculate the sensitivity coefficient $K$, the molecular mass is changed by a small amount $f$ and a shifted line list is created. This slight shift in molecular mass allows us to simulate a variation in $\mu$. After a thorough investigation into the amount to change the molecular mass, we decided on a scale multiplicity of $f = 1.0001$. 
This is chosen to be small enough to ensure there are minimal swapping of energy level assignments within \textsc{Duo}, and large enough to avoid significant numerical noise interference. 
This allows us to calculate $K$ using equation \ref{eqn:K}, where the derivatives are numerically calculated with finite differences. 
To account for the importance of temperature and spectral intensity of transitions in determining their appropriateness as a probe of variation in $\mu$, we choose here to keep only transitions with intensities of greater than $10^{-30}$ cm/molecule at 1000 K. This constraint will retain all transitions of potential interest as all transitions excluded are extremely unlikely to be observable in any astrophysical environment; indeed many transitions we retain are unlikely to be observable as they will be too weak. 

\subsection{Results and Discussion}\label{ssct:res}
We have calculated the sensitivity coefficients for all allowed transitions of the 11 astrophysically relevant diatomic molecules for which we have spectroscopic models for the nuclear motion program \textsc{Duo}.

\begin{figure*}
    \centering
    \subfloat[The 8395312 transitions of the 11 diatomic molecules across all frequency range.]{\includegraphics[width=0.45\textwidth]{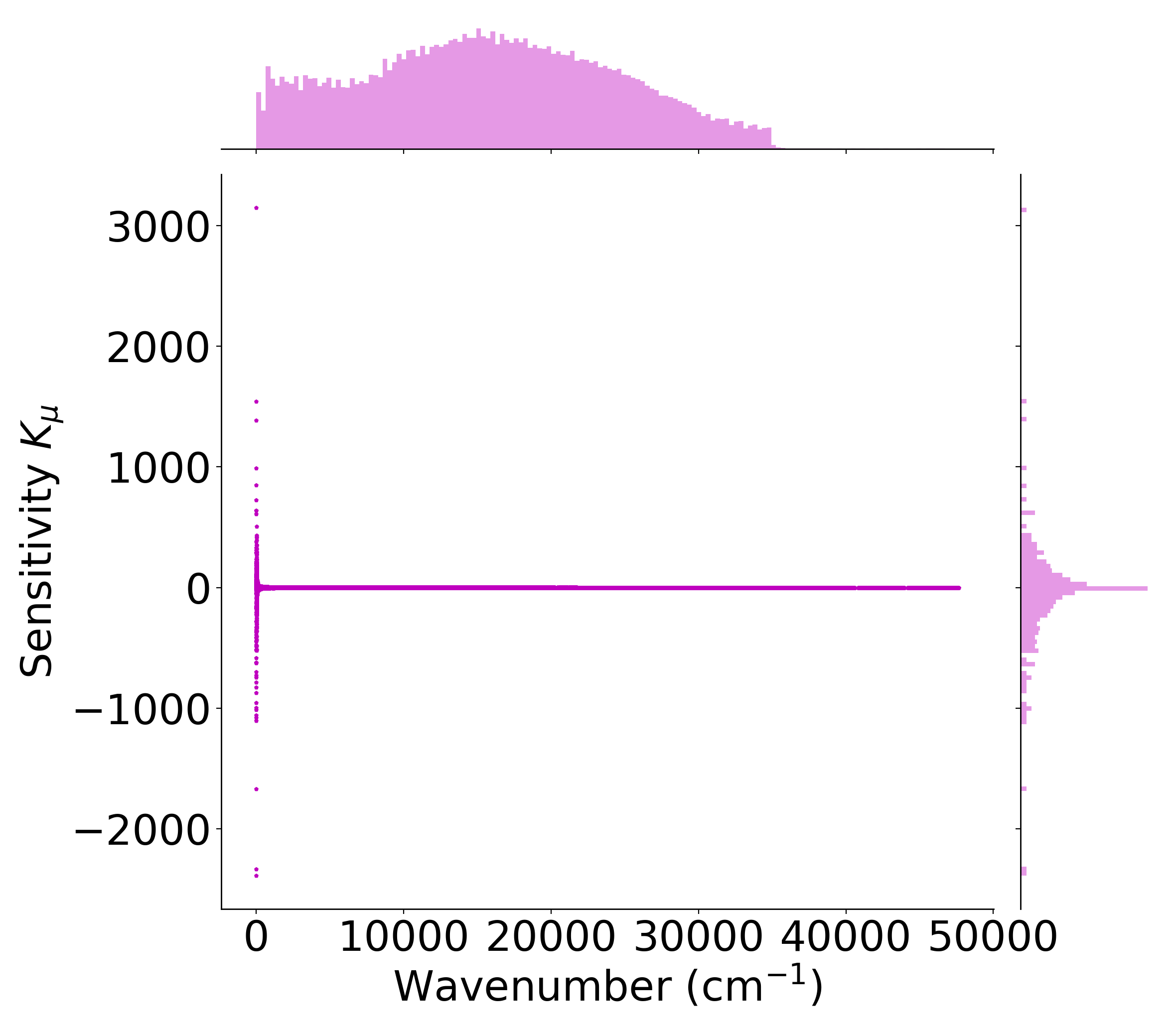}\label{fig:Kwavenumber_all}}
    \qquad
    \subfloat[The 6156371 transitions with transition frequency above 10000cm$^{-1}$, but sensitivity is small.\label{fig:Kwavenumber_g10k}]{\includegraphics[width=0.45\textwidth]{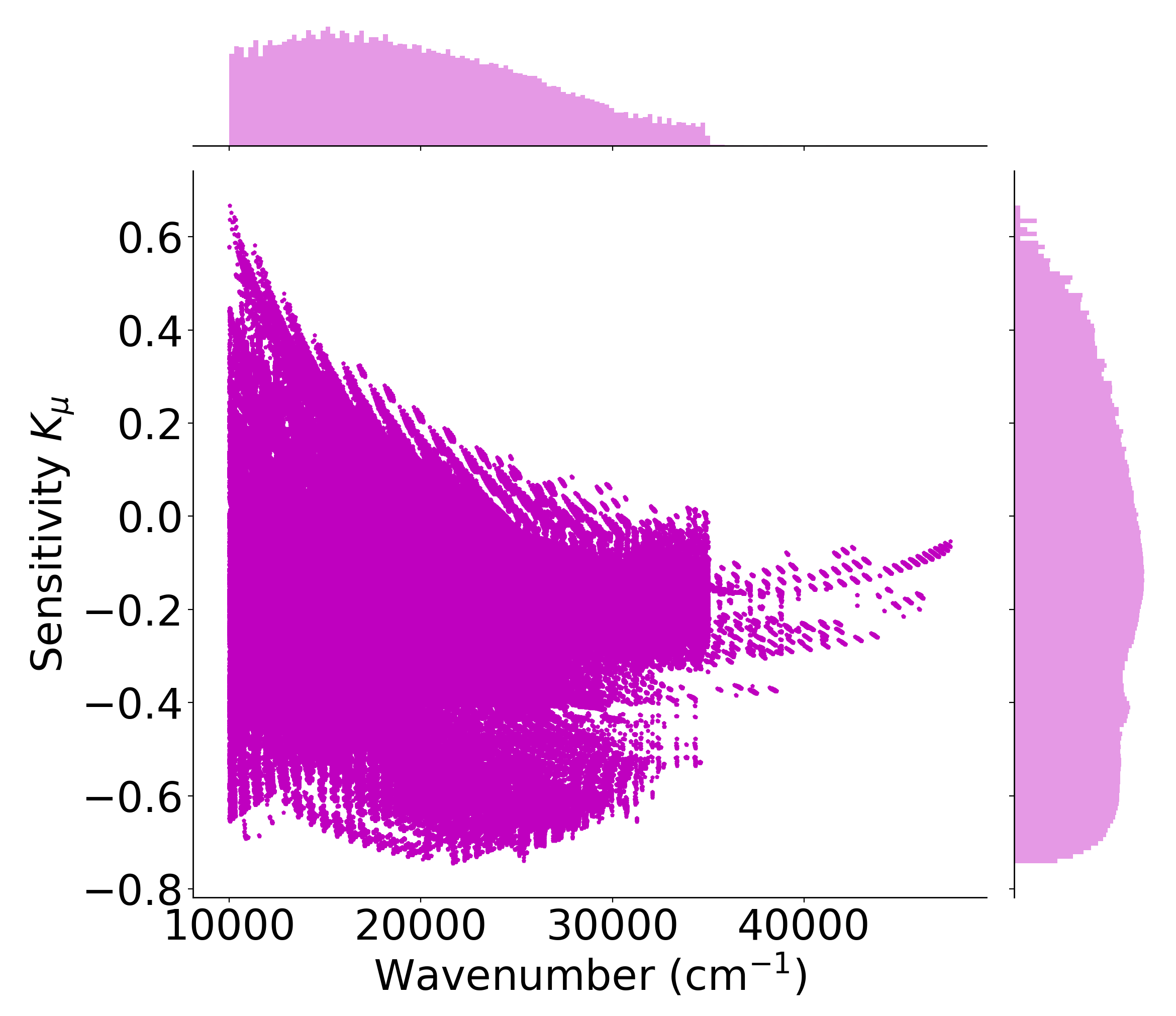}}
    
    \subfloat[The 8199267 transitions with transition frequency above 1000cm$^{-1}$, with increased but not enhanced sensitivity.]{\includegraphics[width=0.45\textwidth]{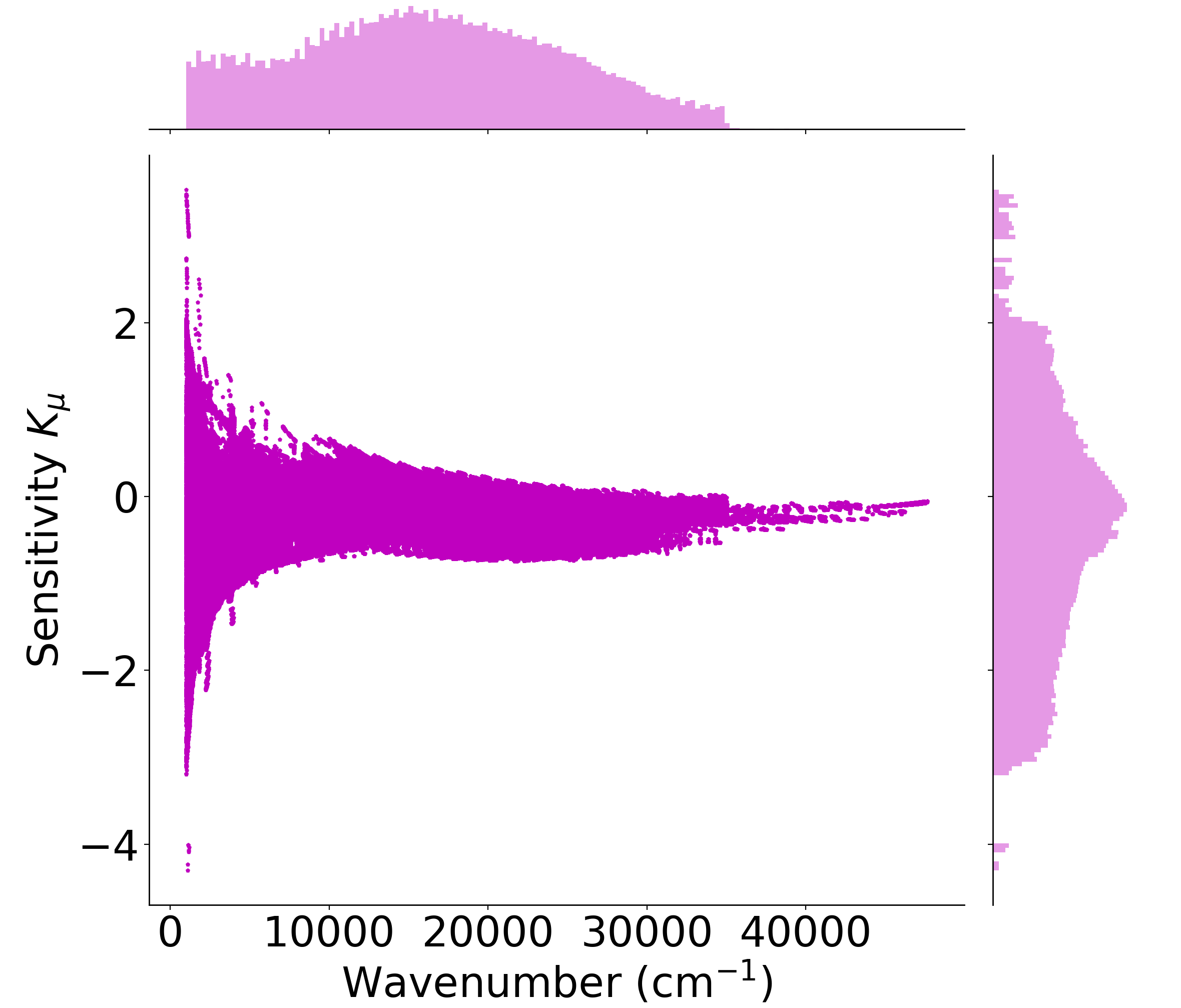}\label{fig:Kwavenumber_g1k}}
    \qquad
    \subfloat[The 196045 transitions with transition frequency below 1000cm$^{-1}$, capped at $|K|<1000$.\label{fig:Kwavenumber_leq1k}]{\includegraphics[width=0.45\textwidth]{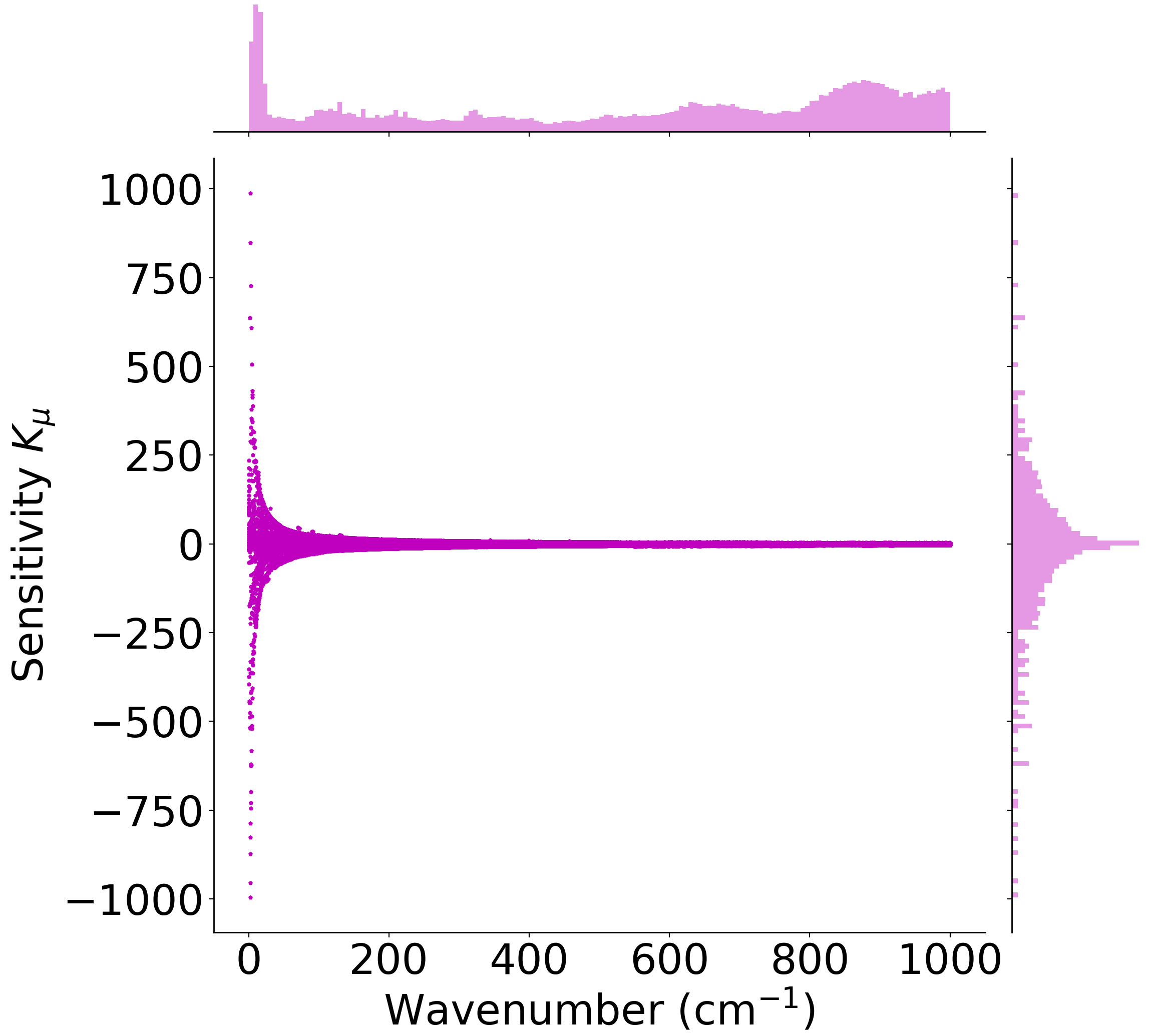}}
    \caption{Distribution of sensitivity coefficients of transitions across various wavenumber regions for all 11 diatomic molecules, with bar plots above each subplot showing the count distribution of the transitions across wavenumber, and to the right showing the count distribution across sensitivity.}\label{fig:dist}
\end{figure*}

\Cref{tbl:SM} shows all the molecules investigated, along with the number of electronic states included in their spectroscopic models, number of transitions produced, number of enhanced transitions, as well as information about the maximum sensitivities. 
From this we can see that \ce{AlO}, \ce{ScH}, \ce{VO} have the largest number of enhanced transitions with intensities greater than $10^{-30}$ cm/molecule at 1000 K, with \ce{C2}, also having a significant number of enhanced transitions. 
The number of enhanced transitions in each diatomic was considered at different temperatures (1000 K, 100 K, and 20 K) with higher intensity cutoffs. Unfortunately for all of the diatomics considered there are no enhanced transitions at temperatures of 100 K and 20 K, and for 1000 K all enhanced transitions have an intensity less than $10^{-22}$ cm/molecule. Nevertheless, we will consider in-depth the factors that cause enhanced transitions in these types of systems in order to predict enhanced  transitions in other astrophysically relevant diatomics, as well as provide insight into understanding enhanced  transitions in diatomics in Earth-based laboratory systems. 

\Cref{fig:Kwavenumber_all} shows the sensitivity coefficients of all of the diatomics investigated as a function of wavenumber. From this, we can see that the distribution of transitions is not particularly concentrated across any wavenumber region. \Cref{fig:Kwavenumber_g10k} examines the distribution of transitions with transition frequency greater than 10,000 \cm. All of these transitions have a sensitivity coefficient between $-0.8 < K < 0.7$. These transitions are not enhanced, and would not be useful primary probes unless the abundance of the molecule was similar to that of \ce{CO} or \ce{H2}. 

We see from \Cref{fig:Kwavenumber_g1k} and \ref{fig:Kwavenumber_leq1k} that all enhanced transitions are all below 1000 \cm, but that not all transitions with a frequency less than 1000 \cm are enhanced. We will thus focus our attention on this region. 

The sensitivity coefficients for these enhanced transitions increases dramatically as the frequency decreases, with enhanced transitions all having frequencies less than 1000 \cm{}. Slightly enhanced rovibronic transitions are likely to have frequencies in the far infrared around 1000 \cm{}, a difficult region for observations even accounting for redshift. Highly enhanced rovibronic transitions may have low frequencies of 10-50 \cm{}, within the range of the ALMA telescope and thus more observable.

In our numerical examples, the existence of more than one electronic state in the spectroscopic model seems to be a necessary, but not sufficient, condition for enhanced transitions. Analysis of sensitivity coefficients in a Morse oscillator support this analysis. 
Given the Morse oscillator potential of 
\begin{equation}
    V(R) = D_e [1 - e^{-\alpha(R-R_e)}]^2, 
    \label{eqn:morsepot}
\end{equation}
where $D_e$ = dissociation energy relative the bottom of the potential, $\alpha = \sqrt{k/2D_e}$, where $k$ is the force constant, $k= (d^2 V/dx^2)_{(x = x_e)}$, and $\nu = \sqrt{\frac{k}{\mu}}$, 
the $n$th vibrational energy level, $E_\textrm{morse}[n]$ are given by
\begin{equation}
    E_\textrm{morse}[n] = h\nu (n + \frac{1}{2}) - \frac{(h\nu)^2}{4D_e}(n+\frac{1}{2})^2.
\end{equation}
By utilising equation \ref{eqn:K} we can calculate the sensitivity coefficients of an anharmonic transition as 
\begin{equation}
    K_{\mu}(i \rightarrow j) =\frac{(-2 D_e + h (1 + i + j) \sqrt{\frac{k}{\mu}})}{(4 D_e - h (1 + i + j) \sqrt{\frac{k}{\mu}})}.
\end{equation}
As $D_e$ goes to infinity the result approaches that of a harmonic oscillator and thus $K \rightarrow -\frac{1}{2}$ as expected. As $D_e$ goes to 0 the anharmonic portion completely dominates and $K \rightarrow -1$. The analysis doesn't substantially change with the addition of rotational energies. Therefore, we expect a single electronic state to not display enhanced transition sensitivities in the absence of fine structure effects.

    

\begin{figure}
    \centering
    \subfloat[\ce{^{51}V^{16}O}]{ \includegraphics[width=0.45\textwidth]{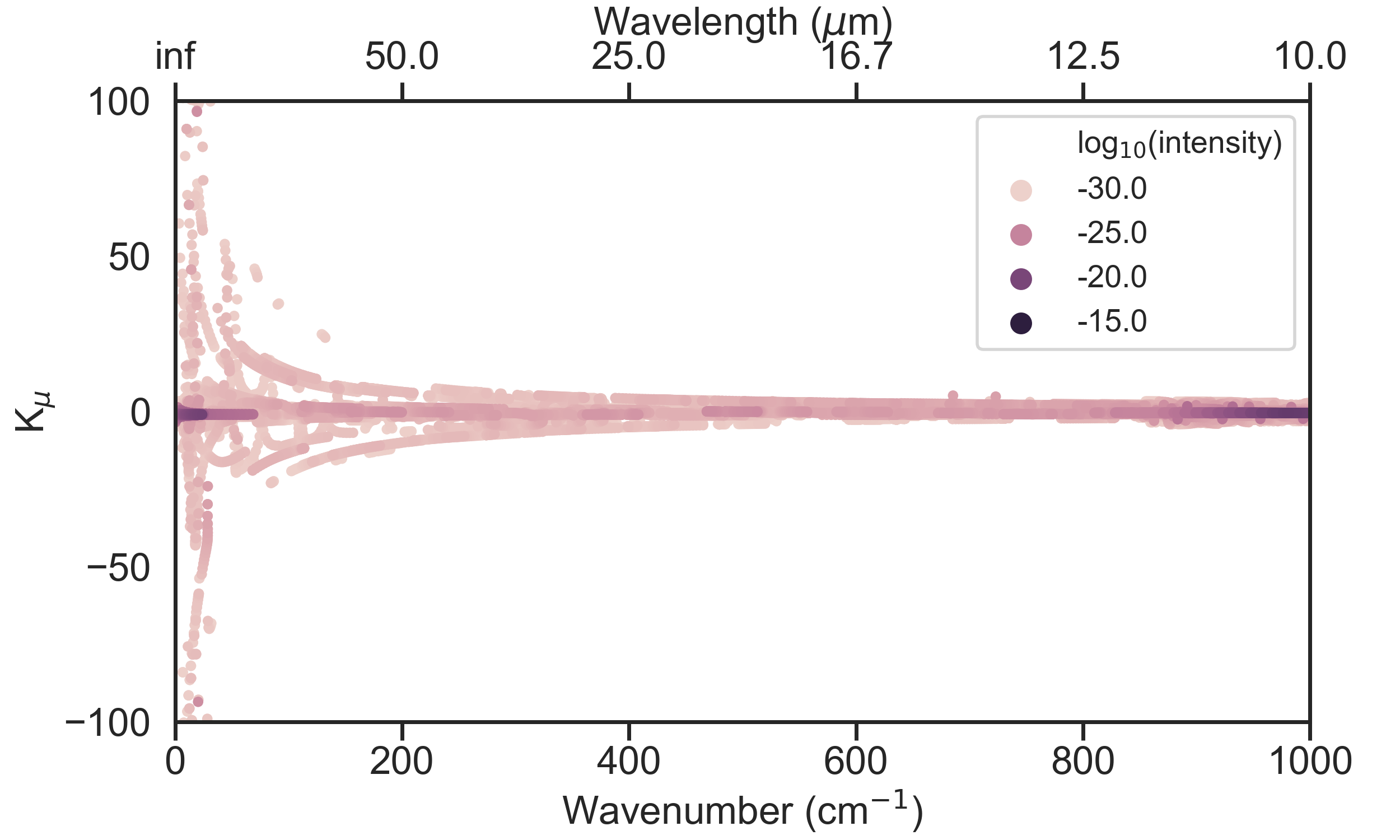}\label{fig:VO}}

    \subfloat[\ce{^{27}Al^{16}O}\label{fig:AlO}]{\includegraphics[width=0.45\textwidth]{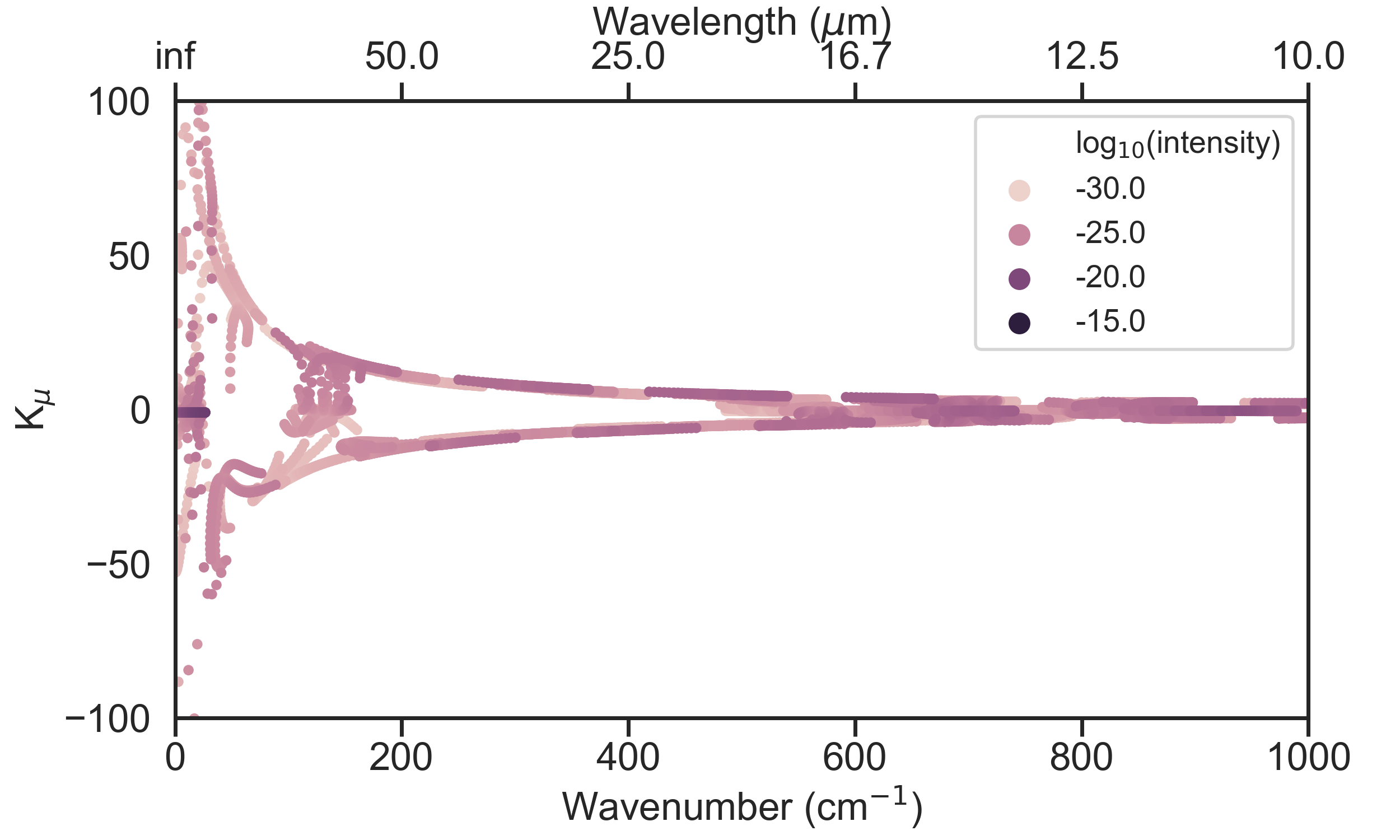}}
    \caption{Distribution of the sensitivities of \ce{^{51}V^{16}O} and \ce{^{27}Al^{16}O} transitions with frequencies below 1000 \cm, capped at sensitivities with magnitude less than 100. The colour shows the intensity of the transitions. }\label{fig:multiES}
\end{figure}

However, the existence of multiple electronic states is clearly not the only important factor; for example, TiO and SiH have almost no enhanced transitions. For the molecules that do show enhanced transitions, we see in \Cref{fig:multiES} for \ce{^{51}V^{16}O} and \ce{^{27}Al^{16}O} that there is a characteristic relationship between sensitivity coefficient $K$ versus wavenumber of the transition with a strong increase in $K$ at small wavenumbers. 

\begin{figure}
    \centering
    \includegraphics[width=0.4\textwidth]{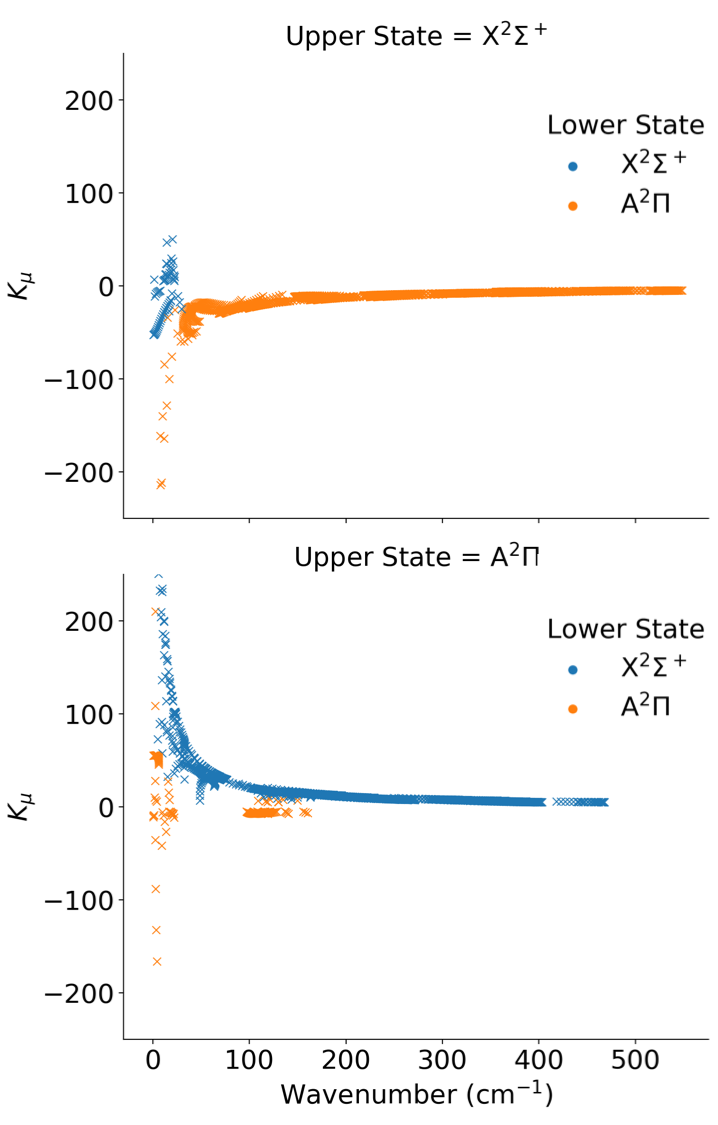}
    \caption{Distribution of the sensitivities of the 3948 enhanced ($|K|>5$) \ce{^{27}Al^{16}O} transitions across wavenumber, where the colour represents the electronic state on the lower energy level of the transition, and where the upper panel shows the transitions with \X{} electronic state as the upper energy level, and the lower panel shows the transitions with \A{} electronic state as the upper energy level}
    \label{fig:AlO_state}
\end{figure}
For rovibronic transitions in diatomic molecules we have found that low frequency transitions that also change electronic states generally have enhanced sensitivity, as shown for AlO in \Cref{fig:AlO_state} where approximately 95\% of enhanced transitions involve a change in electronic state. We propose that the other 5\% are due to coupling between electronic states.

\begin{figure}
    \centering
    \includegraphics[width=0.5\textwidth]{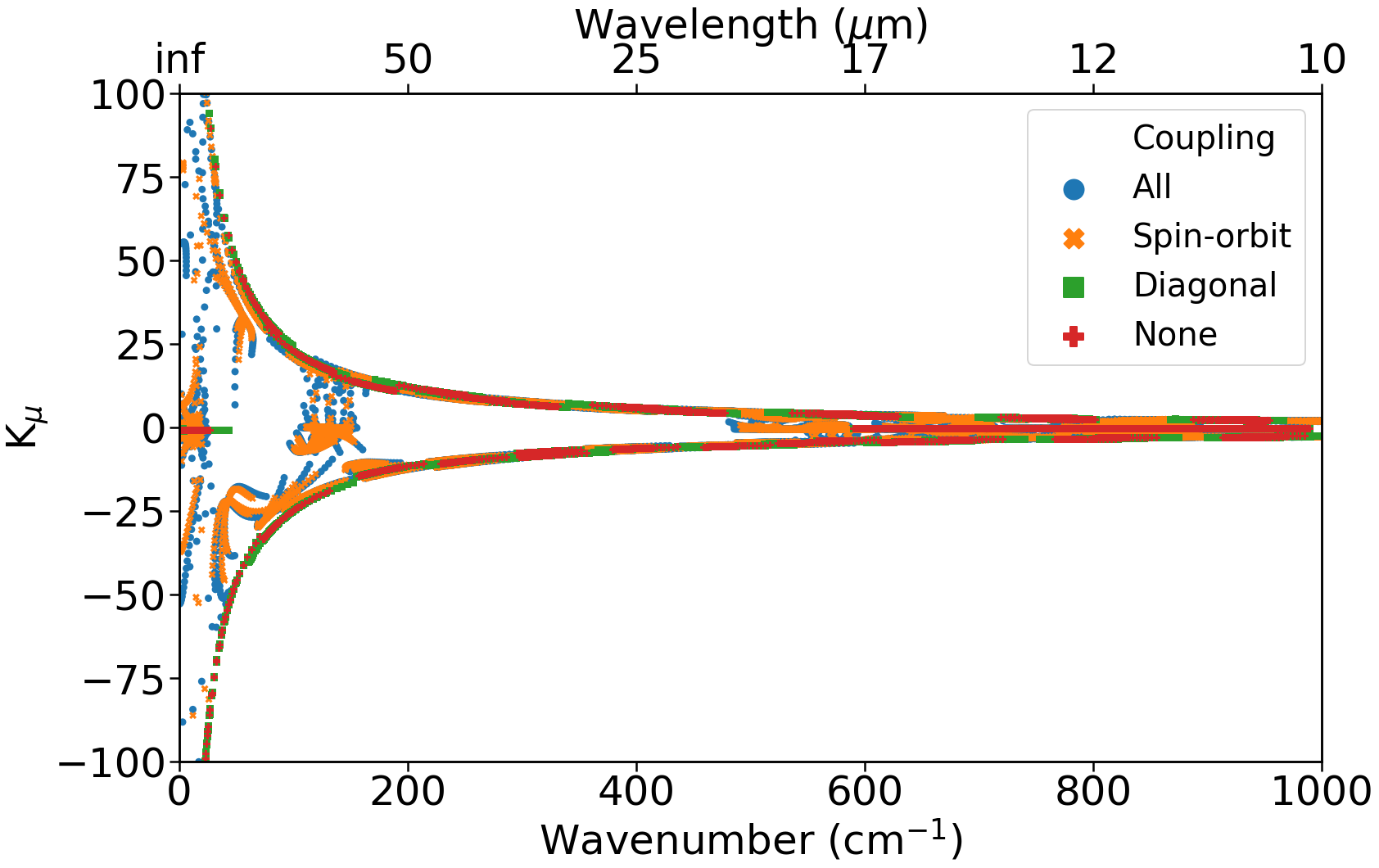}
    \caption{Effects of changing the different coupling within the \ce{AlO} spectroscopic model on the distribution and shape of the sensitivity coefficients across wavenumber.}
    \label{fig:AlO_couplings}
\end{figure}

We can investigate the conditions for enhanced transitions further by modifying the spectroscopic model of AlO, sequentially removing coupling between and within the electronic states.  \Cref{fig:AlO_couplings} shows presents sensitivity coefficients versus wavenumber comparing the full spectroscopic model to the cases where (a) there is only spin-orbit coupling (i.e. spin-spin coupling and electronic angular momentum removed),(b) where there is only diagonal spin-orbit coupling (i.e. removing the \A-\X{} spin-orbit coupling) and (c) where there are no couplings at all. Changing the couplings clearly affects the details of the sensitivity coefficients, but not the overall shape relating sensitivity coefficient to wavenumber. Even with no couplings, there are still clearly strongly enhanced transitions. 

 \begin{figure*}
    \centering
    \includegraphics[width=0.7\textwidth]{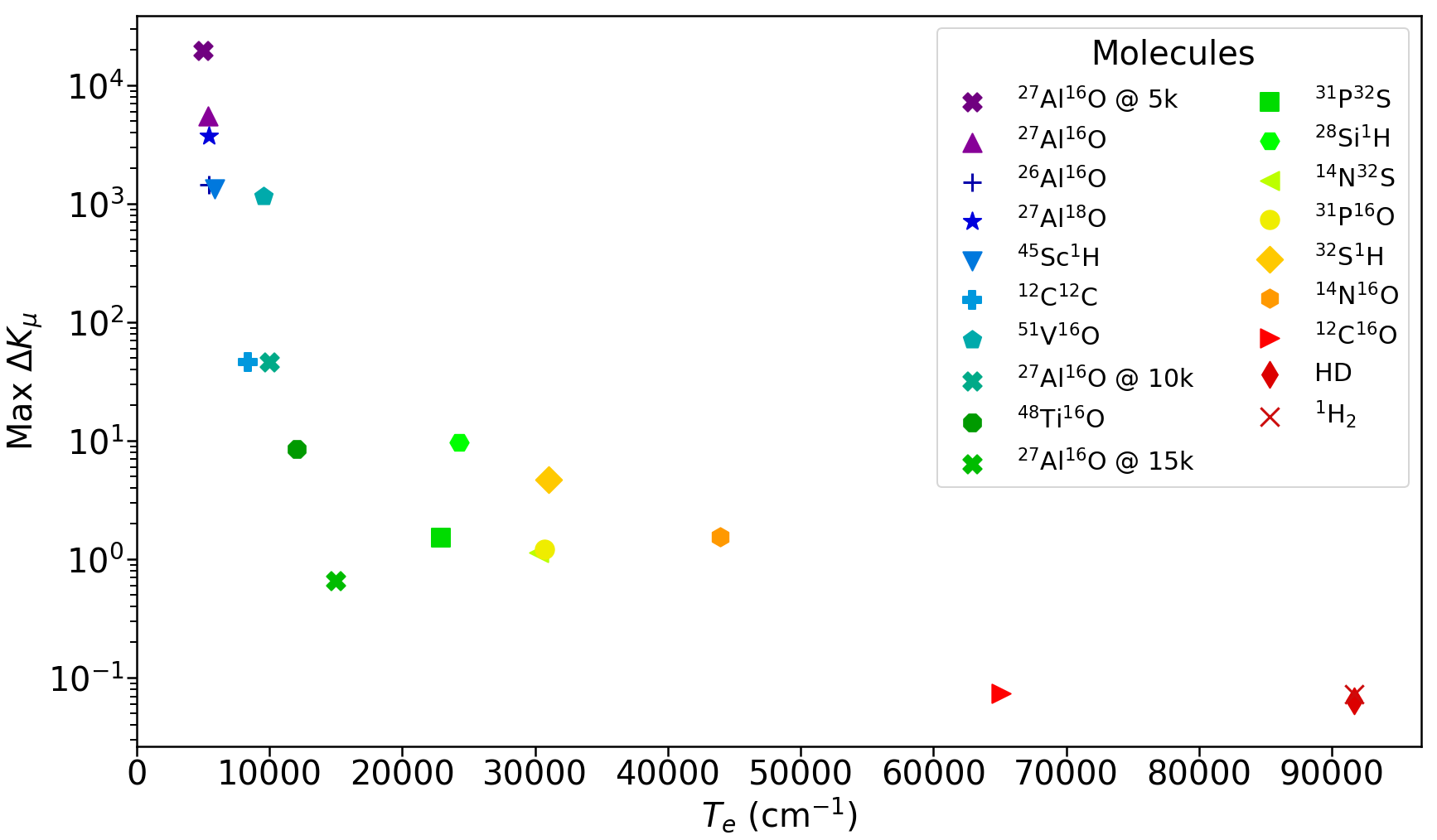}
    \caption{Relation between the energy difference of the ground electronic state and the first spin/symmetry allowed excited electronic state and the maximum $|\Delta K|$ for each diatomic examined within this paper as well as three data point of \ce{^{27}Al^{16}O} where the spectroscopic model was altered by shifting the energy of the \A{} electronic state, denoted with filled crosses. Also included is a combination of the data from tables \ref{tbl:proposed} and \ref{tbl:diatomics} for \ce{CO}, \ce{HD}, and \ce{H2}}
    \label{fig:deltaK_deltaE}
 \end{figure*}

Comparing \Cref{tbl:diatomics} with \Cref{tbl:SM}, we see that if the first electronic state is too high in energy, there are no transitions with enhanced sensitivity; for example, \ce{PS}, \ce{SiH} and \ce{PO} all have $T_e>20000$ \cm. Furthermore, we need to consider selection rules. For example, \ce{TiO} has no enhanced transitions but a low-lying first excited state of \ce{TiO} with a $T_e$ of only 3440 \cm. However, outside of significant mixing (which the specific spin and symmetry of the two states don't allow), transitions between the ground and first electronic states are forbidden. The lowest-lying electronic state with allowed transitions from the ground state has $T_e$ of 12025 \cm{}, accounting for the lack of enhanced transitions in $T_e$.

Conversely, we can see that \ce{AlO}, \ce{ScH}, \ce{C2}, and \ce{VO} all have first allowed electronic states with an energy gap of less than 10,000 \cm and have a substantial number of enhanced transitions. We explore this relationship further in \Cref{fig:deltaK_deltaE} which shows the largest difference in sensitive transitions as a function of the energy gap between the ground state and first allowed excited electronic state. Included in this figure are the three points where we have taken the spectroscopic model for \ce{AlO} and only changed the energy of the first excited electronic state to a $T_e$ of  5,000 \cm, 10,000 \cm, and 15,000 \cm, as well as a combination of data from Tables \ref{tbl:proposed} and \ref{tbl:diatomics} for \ce{H2}, \ce{HD}, and \ce{CO}. 

\Cref{fig:deltaK_deltaE} clearly illustrates that the smaller the $T_e$ of the first allowed electronic excited state, the stronger the enhancement of the sensitivity of the transitions in that molecule. This highlights the lack of sensitive transitions within \ce{CO}, \ce{HD}, and \ce{H2} shown in \Cref{tbl:proposed}. This criterion can be used to help screen molecules as potential astrophysical probes of $\mu$ variation; see \Cref{sct:predict}.


\begin{figure}
    \centering
    \includegraphics[width=0.45\textwidth]{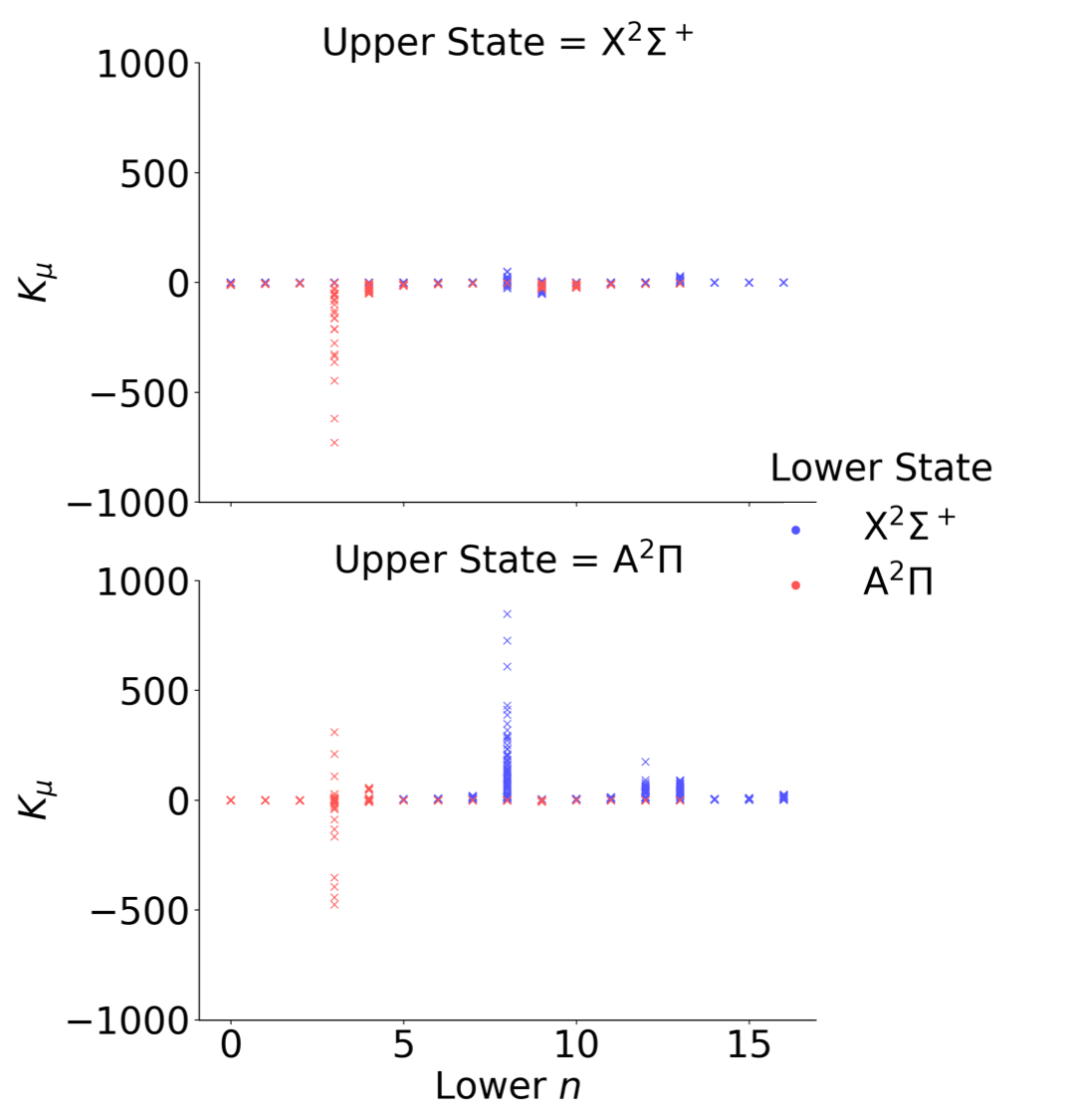}
    \caption{Distribution of sensitivity coefficients across vibrational quantum number $n$ for \ce{AlO}, where the colour represents the electronic state on the lower energy level of the transition, and where the upper panel shows the transitions with \X{} electronic state as the upper energy level, and the lower panel shows the transitions with \A{} electronic state as the upper energy level}
    \label{fig:AlO_state_vib}
\end{figure}

To explore these important variables further, we consider more closely the sensitivity of different types of transitions in AlO.  In \Cref{fig:AlO_state_vib} we can see the distribution of sensitivities across vibrational quantum number, $n$, for \ce{AlO}, split by the electronic state of the upper and lower levels. It is very clear that the bulk of the significantly enhanced transitions involve \X{}($n=8$) and \A{}($n=3$). The enhanced transitions that occur within only the \A{} state are again probably due to mixing between the \X{} and \A{} states, the same transitions we see in \Cref{fig:AlO_state}.

\begin{figure}
    \centering
    \includegraphics[width=0.45\textwidth]{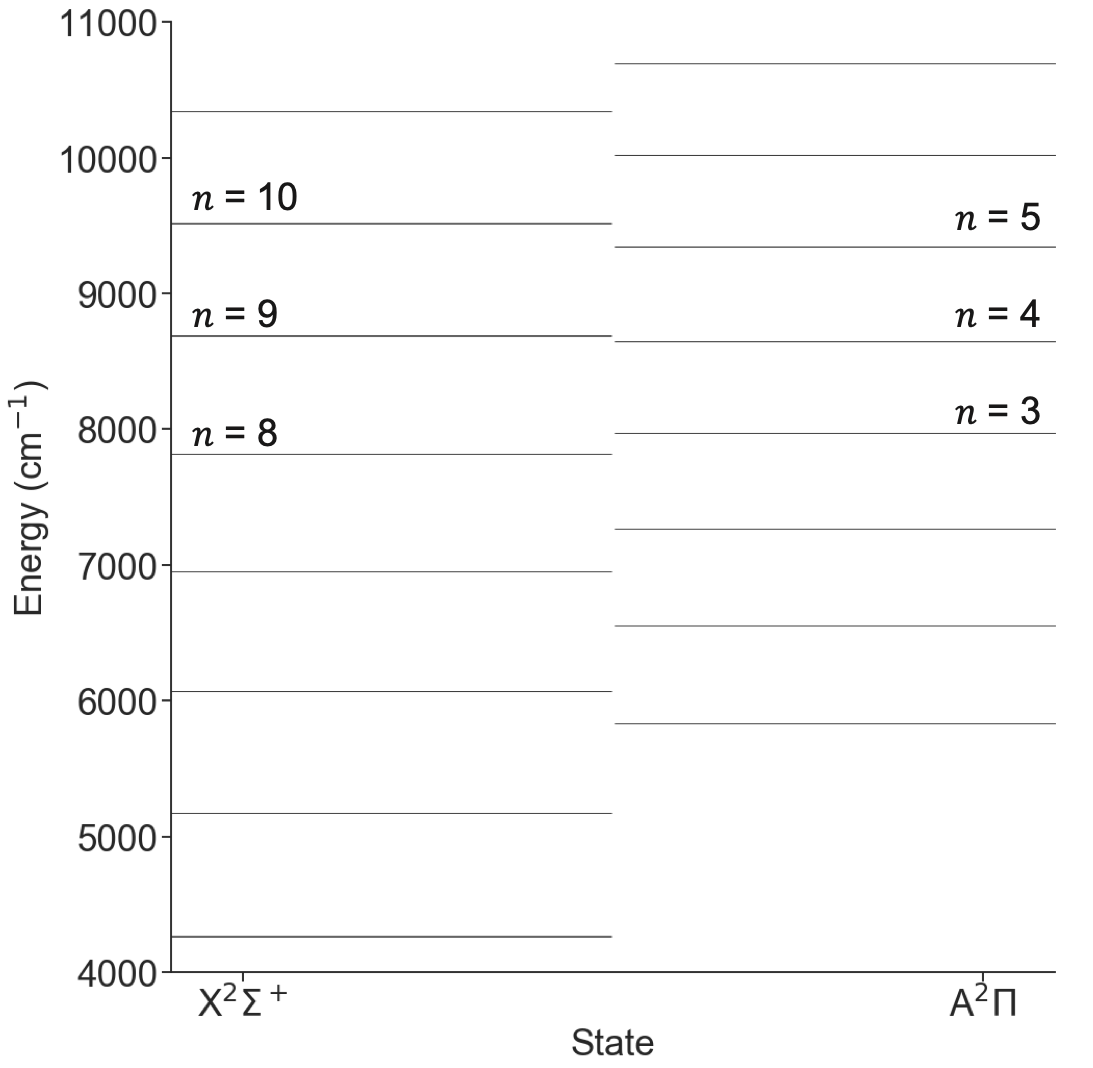}
    \caption{Ground rotational levels of the vibrational energy levels (labelled $n$) within \ce{^{27}Al^{16}O} across the first two electronic energy levels. }
    \label{fig:AlO_energylevels}
\end{figure}

The vibrational energy levels of the \X{} and \A{} states of AlO are shown in \Cref{fig:AlO_energylevels}. It is clear that the \X{}($n=8$) and \A{}($n=3$) are very close, while the \X{}($n=9$) and \A{}($n=4$) levels are near degenerate; this is the cause of the enhanced transitions. In equation \ref{eqn:delta_nu}, a small transition frequency and a regular change in frequency due to variation in $\mu$ will give rise to enhanced sensitivities. 
The population of the \X{}($n=9$) level will be substantially smaller than the population of the \X{}($n=8$) level due to the Boltzmann factor and thus our intensity cut-off has generally removed transitions involving the \X{}($n=9$) level.

\begin{figure}
    \centering
    \includegraphics[width=0.45\textwidth]{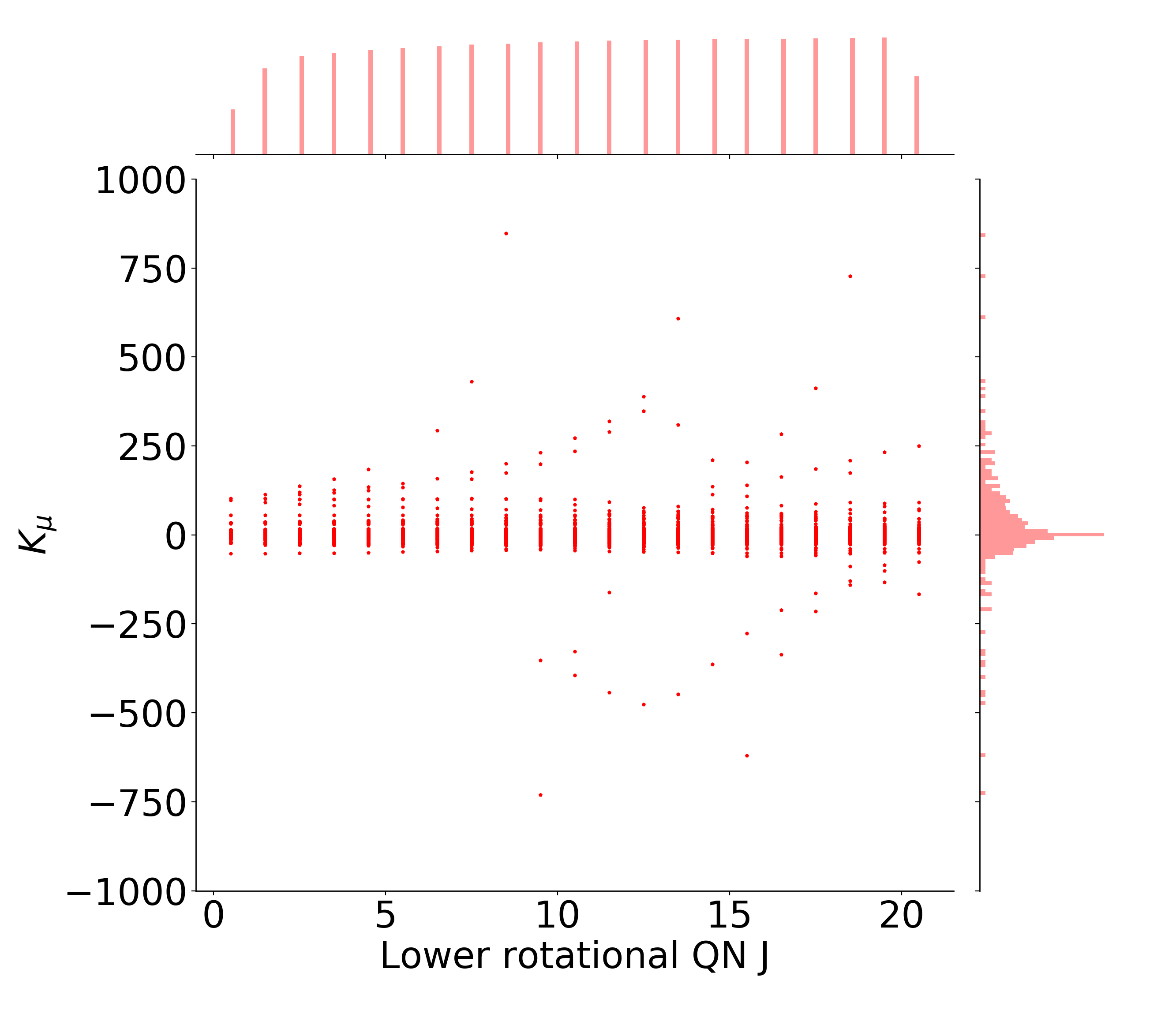}
    \caption{Distribution of sensitivity coefficients across rotational quantum number $J$ for \ce{AlO}, with a bar plot above showing the count distribution of the transitions across lower energy vibrational quantum number $J$, and to the right showing the count distribution across sensitivity.}
    \label{fig:AlO_rot}
\end{figure}

\Cref{fig:AlO_rot} shows the distribution of rotational quantum number and sensitivity for \ce{AlO}. While we can see more of the significantly enhanced, $|K|>1000$ occur at higher J, there are enhanced transitions across all rotational levels. This result shows that the most important factor in causing enhanced transitions in these sorts of systems is the close vibrational levels rather than rotational levels. This criterion should allow us to look at the vibrational band head data alone to determine if a diatomic will be a useful probe to constrain the variation of the proton-to-electron mass ratio, a much easier task than examining the full rovibronic spectroscopy of a molecule.


\section{Screening molecules as probes}\label{sct:predict}
\subsection*{Diatomics} 
\subsubsection*{Enhanced rovibronic transitions}\label{sct:rovib}

From our investigation, we suggest that the sensitivity of rovibronic transitions diatomic molecules to a variation in $\mu$ can be estimated by considering the term energy of first allowed electronic state. We have shown in \Cref{sct:SM}, particularly \Cref{fig:deltaK_deltaE}, that there is a trend between the energy difference between electronic states and the maximum $\Delta K$. 


Molecules with large $T_e$ are associated with very insensitive rovibronic transitions and can thus be excluded from consideration unless their abundance is extremely high (like \ce{H2} and \ce{CO}); in  \Cref{tbl:diatomics} \ce{H2}, \ce{HD}, \ce{HF}, \ce{NO+}, \ce{N2}, and \ce{CO} have $T_e>50,000$ \cm{}. 

More promisingly, the diatomics from \Cref{tbl:diatomics} that have an allowed excited electronic state with energy gap below 10,000 \cm are \ce{AlO}, \ce{SiC}, \ce{CP}, \ce{SiN}, \ce{C2}, and \ce{CN}. These molecules are predicted to have some enhanced transitions, aligning with the results from \Cref{sct:SM} that show enhanced transitions at high temperatures for \ce{AlO} and \ce{C2}. \ce{SiC}, \ce{CP}, \ce{SiN}, and \ce{CN}, however, are strong consideration as probes for $\mu$ variation. Of those 4 diatomic molecules, \ce{CN} is initially the most promising as it has been observed in many different astrophysical environments, including outside of the Milky Way. \cite{13McWaMa.abun} Interestingly, \ce{SiN} has only been observed in circumstellar envelopes, yet has the highest reported relative abundance to \ce{H2}. \cite{03ScLeMe.abun}

\begin{figure}
    \centering
    \includegraphics[width=0.4\textwidth]{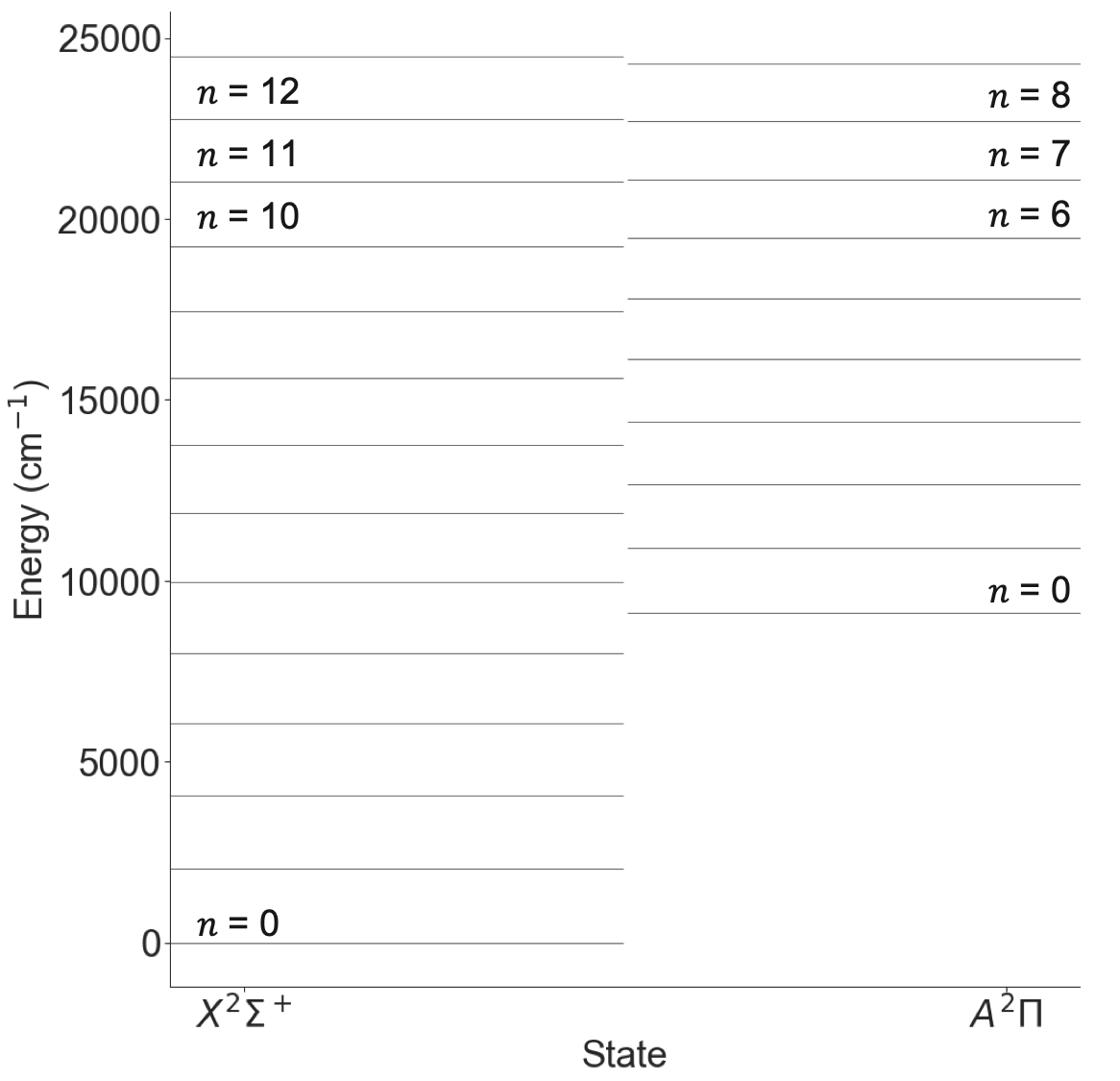}
    \caption{Ground rotational levels of the vibrational energy levels (labelled $n$) within \ce{CN} across the first two electronic energy levels. }
    \label{fig:CN}
\end{figure}

Since there is an available line list for the \ce{CN} radical \cite{14BrRaWe.mp2me} (though no spectroscopic model) we can look at its vibrational band heads to see if any of them are close like we see in \ce{AlO}. \Cref{fig:CN} shows the close-lying vibrational energy levels of \ce{CN}. We can see here that the \X{}($n=11$) and \A{}($n=7$) and the \X{}($n=12$) and \A{}($n=8$) levels are near degenerate. While the energy gap is on the same order of magnitude as the \X{}($n=9$) and \A{}($n=4$) levels in \ce{AlO}, the close-lying vibrational levels in \ce{CN} are much higher in energy and will have a lower Boltzmann population, thus unlikely to be observed. 

\begin{figure}
    \centering
    \includegraphics[width=0.4\textwidth]{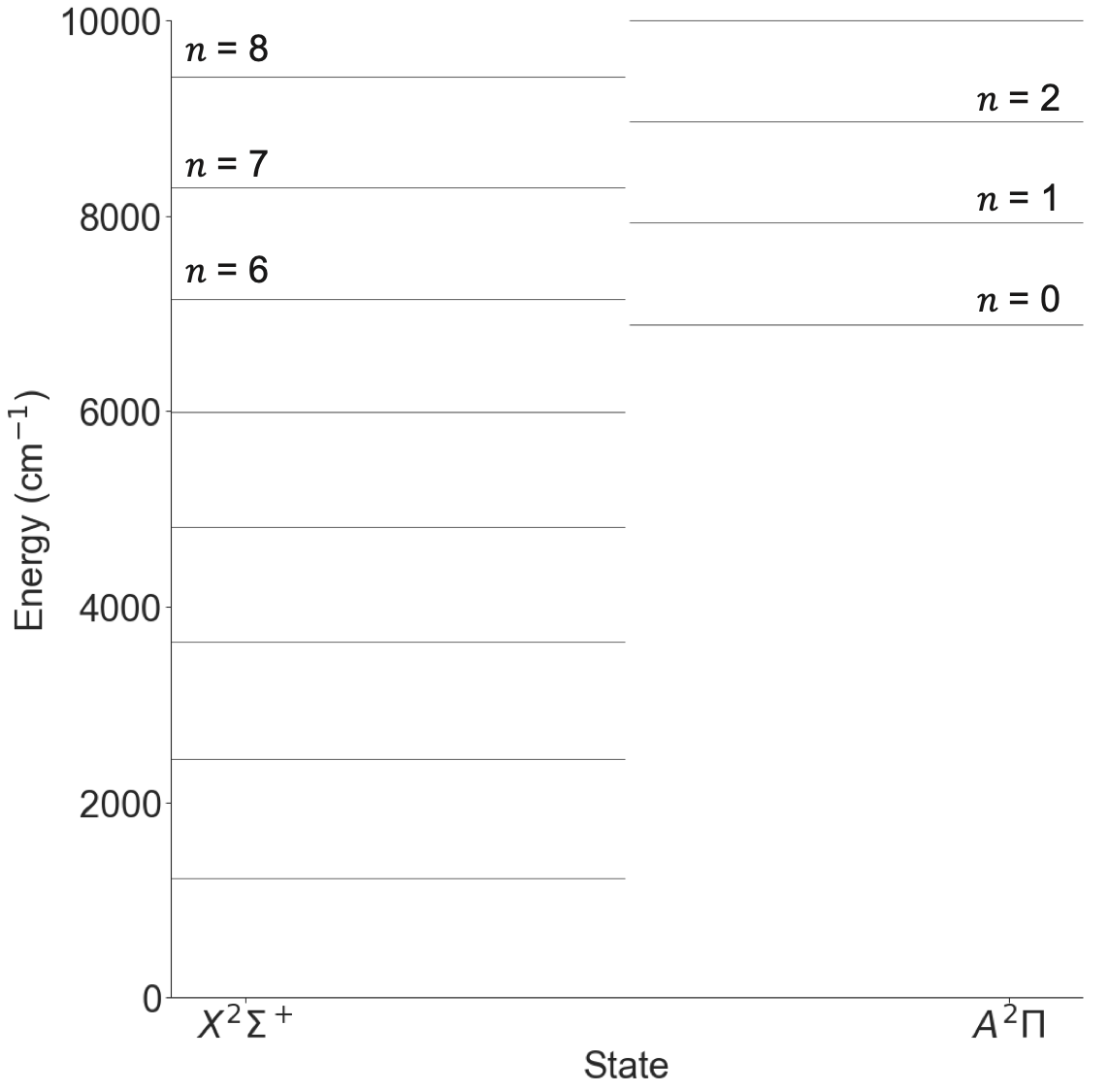}
    \caption{Ground rotational levels of the vibrational energy levels (labelled $n$) within \ce{CP} across the first two electronic energy levels. }
    \label{fig:CP}
\end{figure}

There is a smaller line list available for \ce{CP} as well. \cite{14RaBrWe.mp2me} From the first 8 vibrational bands of each electronic state, none are particularly close, see \Cref{fig:CP}. The \X{}($n=6$) and \A{}($n=0$) states show the most promise, but are almost double the energy gap seen in the \X{}($n=8$) and \A{}($n=3$) states of \ce{AlO}. A more comprehensive line list or spectroscopic model would help in determining the potential of \ce{CP} as a probe. \ce{CP} has only been observed within our Milky Way, with a relative abundance similar to \ce{CN}.



\subsubsection*{Enhanced lambda doubling transitions}\label{sct:lambda}
Previous investigations \cite{09Ko.mp2me,13KoLe.mp2me,13TrHeTo.mp2me, 12DeUbBe.mp2me} have considered transitions involving lambda ($\Lambda$) doubling as potentially useful probes of proton-to-electron mass variation in \ce{NO}, \ce{CH}, \ce{CD}, \ce{CO}, \ce{OH}, and \ce{NH+} and some polyatomics. Lambda doubling occurs when interactions between electronic and rotational motion causes the degeneracy of states with non-zero angular momentum to break. Lambda-doubling shifts the hyperfine energy levels and will shift the sensitivity coefficients. 

Lambda doubling is usually included only empirically into \textsc{Duo}. As such, its mass dependence is not explicitly considered and thus our spectroscopic models cannot in these cases predict how a variation in $\mu$ would be affected by lambda doubling. 

A relevant caveat that may affect published ExoMol line lists slightly is that lambda doubling predicted for non-dominant isotopologues may not be accurate as their values are inherited from the dominant isotopologue of the molecule; this will be considered further in a future publication.

Our tables show that \ce{HCl+}, \ce{SiC}, \ce{SiH} and \ce{SO+} have $\Pi$ ground states but have not yet been considered as probes of $\mu$ variation. Lambda doubling transitions have been observed at medium redshift for \ce{NO} and \ce{SO+} \cite{11MuBeGu.mp2me} as well as \ce{OH} \cite{11Ch.mp2me}. Galactic observations of lambda-doubling transitions of \ce{HCl+} \cite{18Mc.mp2me} suggest possible future interest.

\ce{CN}, \ce{AlO}, \ce{CP}, \ce{C2}, and \ce{SiN} have relatively low ($<$10000 \cm) $\Pi$ electronically excited states, which may be of interest for lambda-doubling. \ce{CN} especially could be interesting as it has been observed in external galaxies. \cite{18Mc.mp2me}

\subsubsection*{Abundant extragalactic molecules}
Molecules with a higher abundance allow for possible use as a probe to constrain the variation of the proton-to-electron mass ratio, even without high sensitive transitions. We already see this with \ce{H2} and \ce{CO}, and to some extent \ce{HD}. From \Cref{tbl:diatomics} we can see that not very many diatomics have an abundance particularly close to \ce{H2}, however, there are a couple that are close to that observed of \ce{CO}. Of the three diatomic (\ce{PO}, \ce{SiS}, \ce{SiO}) with an approximate galactic abundance within an order of magnitude of \ce{CO} only \ce{SiO} has been observed extragalactically. \ce{SiO} relative abundance to \ce{H2} in extragalactic environments is significantly lower, more on the order of $10^{-10}$. \cite{11MuBeGu.mp2me} From this, it is unlikely that any other diatomics have a high enough abundance to be particularly useful probes without enhanced sensitivities. 

\subsection*{Predictions for Polyatomic Molecules}
Our results for diatomic molecules suggest that looking for near degeneracy in vibrational origins of different vibrational modes alone may be sufficient information to screen potential polyatomic molecules for potentially enhanced transitions without the need for full rovibrational calculations. Future work will test this hypothesis using 7 available polyatomic line lists and spectroscopic models: \ce{H2O}, \ce{HCN}, \ce{H2S}, \ce{SO2}, and \ce{H3+} will use the triatomic nuclear motion program DVR3D,\cite{DVR3D} and \ce{H2CO} and \ce{NH3} will use the polyatomic nuclear motion program \textsc{Trove}.\cite{TROVE}  

\section{Conclusions}\label{sct:con}
Here, we look for new molecular probes of proton-to-electron mass variation, $\mu$, over cosmological time, highlighting the importance of considering both the accuracy with which you can measure the transition and the sensitivity of the transition to variation in $\mu$. 

Astrophysical observability of the spectral transition is key to a useful suggestion of a potential probe for $\mu$ variation over cosmological time. To assist chemists, we collate relevant information here.  
In particular, we provide the abundances and relevant physical properties of all known interstellar and extragalactic diatomics and identify all extragalatically observed polyatomic molecules. Further, we discuss the importance of considering the frequency range and spectral resolution of available telescopes to measure a proposed transition. Finally, we emphasis the importance of spectral intensity as a metric for quantifying the suitability of a particular transition. 

For the first time, rovibronic transitions in 11 diatomic molecules are considered as potential probes for $\mu$ variation using high level variational spectroscopic models that include complex coupling between electronic states and transition intensity information. While none of the 11 diatomics investigated have enhanced transitions likely to be observable extragalactically in astrophysical temperatures, we could study the factors leading to sensitive transitions by considering lower intensity transitions that occured at higher temperatures. At 1000 K, low intensity enhanced transitions were observed at low frequencies due to near-degeneracy of energy levels of different kinds.  Within the limitations of the spectroscopic models available, enhanced transitions were only observed due to the degeneracy between vibrational levels of a ground and excited electronic state. Our results showed no enhanced transitions in $^2\Pi$ ground states due to lambda doubling; the empirical manner by which lambda doubling was included into existing spectroscopic models may have limited the ability of this study to identify any enhanced lambda doubling transitions. 

Based on these results, we used the presence of intense transitions to low-lying electronic state (approximately <10,000 \cm) as an initial screening tool for enhanced sensitivity in rovibronic transitions in diatomic molecules, with the next step to investigate the vibrational band heads.  The molecules \ce{SiC}, \ce{CP}, \ce{SiN}, and \ce{CN} seem most promising for further investigation. 

Of the 43 known interstellar diatomic molecules, we identified
\begin{itemize}
    \item 6 molecules (\ce{AlO}, \ce{SiC}, \ce{CP}, \ce{SiN}, \ce{C2}, \ce{CN}) as having potential to possess enhanced rovibronic transitions specifically due to low-frequency transitions between low-lying electronic states;
    \item 10 molecules (\ce{SO+}, \ce{PO}, \ce{OH}, \ce{CH}, \ce{SH}, \ce{NO}, \ce{NS}, \ce{HCl+}, \ce{SiC}, \ce{SiH}) of interest in investigations of sensitive lambda doubling transitions, with \ce{PO} of particular interest due to its high galactic abundances;
    \item 2 molecules (\ce{SiS}, \ce{SiO}) with galactic abundances similar to \ce{CO} which may be of interest for further study (like \ce{H2} and \ce{CO} have been) despite their predicted low sensitivity coefficients from either lambda doubling or rovibronic transitions;
    \item 14 molecules  (\ce{HD}, \ce{HF}, \ce{NO+ }, \ce{N2}, \ce{PN}, \ce{CS}, \ce{SO}, \ce{AlCl}, \ce{O2}, \ce{NH}, \ce{OH+}, \ce{CH+}, \ce{CO+}, \ce{FeO}) that have a low astrophysical abundance, and 11 other molecules with unknown astrophysical abundance  ( \ce{TiO}, \ce{SiH}, \ce{AlF}, \ce{KCl}, \ce{CF+}, \ce{HCl}, \ce{NS+}, \ce{ArH+}, \ce{NaCl}, \ce{SH+}, \ce{CN$^-$}) with low potential for transitions for enhanced sensitivity.
\end{itemize}




Our results also have implications for the use of rovibronic diatomic transitions as probes for variation of $\mu$ in other situations. Astrophysically, one may explore spatial dependence or variation within high gravitational fields. \cite{16UbKoEi.mp2me}  Earth-based experiments can utilise a much larger range of potential molecules\cite{15PaBoFl.mp2me} and also overcome the Boltzmann factor by using multiple lasers to first populate a high-lying initial state, then to drive the desired transition. For example, rovibronic transitions in KRb have recently\cite{19KoOgIn.mpme} been used to constrain $\mu$ variation to a fractional change of less than $10^{-14}$/yr. 



\section*{Acknowledgements} This research was undertaken with the assistance of resources from the National Computational Infrastructure (NCI Australia), an NCRIS enabled capability supported by the Australian Government.

The authors declare no conflicts of interest. 


\bibliography{Muology_paper1}

\begin{thebibliography}{100}

\bibitem{08Al.mp2me}
{\sc M.~Alcubierre},
\newblock {\em {Introduction to 3+1 Numerical Relativity}}, volume~21,
\newblock Oxford University Press, 2008.

\bibitem{17Ra.mp2me}
{\sc S.~Raby},
\newblock {\em {Supersymmetric Grand Unified Theories}}, volume 939 of {\em
  Lecture Notes in Physics},
\newblock Springer International Publishing, Cham, 2017.

\bibitem{08MuFlMu.mp2me}
{\sc M.~T. Murphy}, {\sc V.~V. Flambaum}, {\sc S.~Muller}, and {\sc C.~Henkel},
\newblock {\em Science (New York, N.Y.)} {\bf 320}, 1611 (2008).

\bibitem{11Uz.mp2me}
{\sc J.~P. Uzan},
\newblock {\em Living Reviews in Relativity} {\bf 14}, 2 (2011).

\bibitem{18SaBuDe.mp2me}
{\sc M.~S. Safronova}, {\sc D.~Budker}, {\sc D.~DeMille}, {\sc D.~F.~J.
  Kimball}, {\sc A.~Derevianko}, and {\sc C.~W. Clark},
\newblock {\em Reviews of Modern Physics} {\bf 90}, 025008 (2018).

\bibitem{15Uz.mp2me}
{\sc J.~P. Uzan},
\newblock {\em Comptes Rendus Physique} {\bf 16}, 576 (2015).

\bibitem{14JaBeUb.mp2me}
{\sc P.~Jansen}, {\sc H.~L. Bethlem}, and {\sc W.~Ubachs},
\newblock {\em Journal of Chemical Physics} {\bf 140}, 1 (2014).

\bibitem{16UbBaSa.mp2me}
{\sc W.~Ubachs}, {\sc J.~Bagdonaite}, {\sc E.~J. Salumbides}, {\sc M.~T.
  Murphy}, and {\sc L.~Kaper},
\newblock {\em Reviews of Modern Physics} {\bf 88}, 1 (2016).

\bibitem{11Th.mp2me}
{\sc R.~I. Thompson},
\newblock {Current State of mp/me= {\$}{$\mu$}{\$} Measurements Versus Cosmic
  Time},
\newblock in {\em From Varying Couplings to Fundamental Physics}, edited by
  {\sc C.~Martins} and {\sc P.~Molaro}, pp. 77--87, Berlin, Heidelberg, 2011,
  Springer Berlin Heidelberg.

\bibitem{17Ma.mp2me}
{\sc C.~Martins},
\newblock {\em Reports on Progress in Physics} {\bf 80}, 126902 (2017).

\bibitem{99WeFlCh.mp2me}
{\sc J.~K. Webb}, {\sc V.~V. Flambaum}, {\sc C.~W. Churchill}, {\sc M.~J.
  Drinkwater}, and {\sc J.~D. Barrow},
\newblock {\em Physical Review Letters} {\bf 82}, 884 (1999).

\bibitem{11JaXuKl.mp2me}
{\sc P.~Jansen}, {\sc L.~Xu}, {\sc I.~Kleiner}, {\sc W.~Ubachs}, and {\sc H.~L.
  Bethlem},
\newblock {\em Physical Review Letters} {\bf 106}, 100801 (2011).

\bibitem{08ShBuCh.mp2me}
{\sc A.~Shelkovnikov}, {\sc R.~J. Butcher}, {\sc C.~Chardonnet}, and {\sc
  A.~Amy-Klein},
\newblock {\em Physical Review Letters} {\bf 100}, 150801 (2008).

\bibitem{19KoOgIn.mpme}
{\sc J.~Kobayashi}, {\sc A.~Ogino}, and {\sc S.~Inouye},
\newblock {\em Nature communications} {\bf 10}, 1 (2019).

\bibitem{75Th.mp2me}
{\sc R.~I. Thompson},
\newblock {\em Astrophysical Letters} {\bf 16}, 3 (1975).

\bibitem{11BeBoFl.mp2me}
{\sc K.~Beloy}, {\sc A.~Borschevsky}, {\sc V.~V. Flambaum}, and {\sc
  P.~Schwerdtfeger},
\newblock {\em Physical Review A} {\bf 84}, 1 (2011).

\bibitem{13KoLe.mp2me}
{\sc M.~G. Kozlov} and {\sc S.~A. Levshakov},
\newblock {\em Annalen der Physik} {\bf 525}, 452 (2013).

\bibitem{11LeKoRe.mp2me}
{\sc S.~A. Levshakov}, {\sc M.~G. Kozlov}, and {\sc D.~Reimers},
\newblock {\em The Astrophysical Journal} {\bf 738}, 26 (2011).

\bibitem{13Ko.mp2me}
{\sc M.~G. Kozlov},
\newblock {\em Physical Review A} {\bf 87}, 1 (2013).

\bibitem{09Ko.mp2me}
{\sc M.~G. Kozlov},
\newblock {\em Physical Review A} {\bf 80}, 1 (2009).

\bibitem{12IlJaKo.mp2me}
{\sc V.~V. Ilyushin}, {\sc P.~Jansen}, {\sc M.~G. Kozlov}, {\sc S.~A.
  Levshakov}, {\sc I.~Kleiner}, {\sc W.~Ubachs}, and {\sc H.~L. Bethlem},
\newblock {\em Physical Review A} {\bf 85}, 032505 (2012).

\bibitem{15OwYuPo.mp2me}
{\sc A.~Owens}, {\sc S.~N. Yurchenko}, {\sc O.~L. Polyansky}, {\sc R.~I.
  Ovsyannikov}, {\sc W.~Thiel}, and {\sc V.~{\v{S}}pirko},
\newblock {\em Monthly Notices of the Royal Astronomical Society} {\bf 454},
  2292 (2015).

\bibitem{18OwYuSp.mp2me}
{\sc A.~Owens}, {\sc S.~N. Yurchenko}, and {\sc V.~{\v{S}}pirko},
\newblock {\em Monthly Notices of the Royal Astronomical Society} {\bf 473},
  4986 (2018).

\bibitem{12KaLaSt.mp2me}
{\sc N.~Kanekar}, {\sc G.~I. Langston}, {\sc J.~T. Stocke}, {\sc C.~L.
  Carilli}, and {\sc K.~M. Menten},
\newblock {\em The Astrophysical Journal} {\bf 746}, L16 (2012).

\bibitem{15BaUbMu.mp2me}
{\sc J.~Bagdonaite}, {\sc W.~Ubachs}, {\sc M.~T. Murphy}, and {\sc J.~B.
  Whitmore},
\newblock {\em Physical Review Letters} {\bf 114}, 071301 (2015).

\bibitem{11WeMo.mp2me}
{\sc M.~Wendt} and {\sc P.~Molaro},
\newblock {\em Astronomy {\&} Astrophysics} {\bf 526}, 1 (2011).

\bibitem{12WeMo.mp2me}
{\sc M.~Wendt} and {\sc P.~Molaro},
\newblock {\em Astronomy {\&} Astrophysics} {\bf 541}, 1 (2012).

\bibitem{08KiWeMu.mp2me}
{\sc J.~A. King}, {\sc J.~K. Webb}, {\sc M.~T. Murphy}, and {\sc R.~F.
  Carswell},
\newblock {\em Physical Review Letters} {\bf 101}, 1 (2008).

\bibitem{08WeRe.mp2me}
{\sc M.~Wendt} and {\sc D.~Reimers},
\newblock {\em The European Physical Journal Special Topics} {\bf 163}, 197
  (2008).

\bibitem{15DaBaMu.mp2me}
{\sc M.~Dapr{\`{a}}}, {\sc J.~Bagdonaite}, {\sc M.~T. Murphy}, and {\sc
  W.~Ubachs},
\newblock {\em Monthly Notices of the Royal Astronomical Society} {\bf 454},
  489 (2015).

\bibitem{14VaRaNo.mp2me}
{\sc D.~Albornoz~V{\'{a}}squez}, {\sc H.~Rahmani}, {\sc P.~Noterdaeme}, {\sc
  P.~Petitjean}, {\sc R.~Srianand}, and {\sc C.~Ledoux},
\newblock {\em Astronomy {\&} Astrophysics} {\bf 562}, A88 (2014).

\bibitem{09ThBeBl.mp2me}
{\sc R.~I. Thompson}, {\sc J.~Bechtold}, {\sc J.~H. Black}, {\sc
  D.~Eisenstein}, {\sc X.~Fan}, {\sc R.~C. Kennicutt}, {\sc C.~Martins}, {\sc
  J.~X. Prochaska}, and {\sc Y.~L. Shirley},
\newblock {\em The Astrophysical Journal} {\bf 703}, 1648 (2009).

\bibitem{07UbBuEi.mp2me}
{\sc W.~Ubachs}, {\sc R.~Buning}, {\sc K.~S.~E. Eikema}, and {\sc E.~Reinhold},
\newblock {\em Journal of Molecular Spectroscopy} {\bf 241}, 155 (2007).

\bibitem{12BaMuKa.mp2me}
{\sc J.~Bagdonaite}, {\sc M.~T. Murphy}, {\sc L.~Kaper}, and {\sc W.~Ubachs},
\newblock {\em Monthly Notices of the Royal Astronomical Society} {\bf 421},
  419 (2012).

\bibitem{13RaWeSr.mp2me}
{\sc H.~Rahmani}, {\sc M.~Wendt}, {\sc R.~Srianand}, {\sc P.~Noterdaeme}, {\sc
  P.~Petitjean}, {\sc P.~Molaro}, {\sc J.~B. Whitmore}, {\sc M.~T. Murphy},
  {\sc M.~Centurion}, {\sc H.~Fathivavsari}, {\sc S.~D'Odorico}, {\sc T.~M.
  Evans}, {\sc S.~A. Levshakov}, {\sc S.~Lopez}, {\sc C.~Martins}, {\sc
  D.~Reimers}, and {\sc G.~Vladilo},
\newblock {\em Monthly Notices of the Royal Astronomical Society} {\bf 435},
  861 (2013).

\bibitem{11KiMuUb.mp2me}
{\sc J.~A. King}, {\sc M.~T. Murphy}, {\sc W.~Ubachs}, and {\sc J.~K. Webb},
\newblock {\em Monthly Notices of the Royal Astronomical Society} {\bf 417},
  3010 (2011).

\bibitem{17DaVaMu.mp2me}
{\sc M.~Dapr{\`{a}}}, {\sc M.~van~der Laan}, {\sc M.~T. Murphy}, and {\sc
  W.~Ubachs},
\newblock {\em Monthly Notices of the Royal Astronomical Society} {\bf 465},
  4057 (2017).

\bibitem{10MaBuMu.mp2me}
{\sc A.~L. Malec}, {\sc R.~Buning}, {\sc M.~T. Murphy}, {\sc N.~Milutinovic},
  {\sc S.~L. Ellison}, {\sc J.~X. Prochaska}, {\sc L.~Kaper}, {\sc
  J.~Tumlinson}, {\sc R.~F. Carswell}, and {\sc W.~Ubachs},
\newblock {\em Monthly Notices of the Royal Astronomical Society} {\bf 403},
  1541 (2010).

\bibitem{11VaMuMa.mp2me}
{\sc F.~van Weerdenburg}, {\sc M.~T. Murphy}, {\sc A.~L. Malec}, {\sc
  L.~Kaper}, and {\sc W.~Ubachs},
\newblock {\em Physical Review Letters} {\bf 106}, 180802 (2011).

\bibitem{16DaNiSa.mp2me}
{\sc M.~Dapr{\`{a}}}, {\sc M.~L. Niu}, {\sc E.~J. Salumbides}, {\sc M.~T.
  Murphy}, and {\sc W.~Ubachs},
\newblock {\em The Astrophysical Journal} {\bf 826}, 192 (2016).

\bibitem{17DaNoVo.mp2me}
{\sc M.~Dapr{\`{a}}}, {\sc P.~Noterdaeme}, {\sc M.~Vonk}, {\sc M.~T. Murphy},
  and {\sc W.~Ubachs},
\newblock {\em Monthly Notices of the Royal Astronomical Society} {\bf 467},
  3848 (2017).

\bibitem{09HeMeMu.mp2me}
{\sc C.~Henkel}, {\sc K.~M. Menten}, {\sc M.~T. Murphy}, {\sc N.~Jethava}, {\sc
  V.~V. Flambaum}, {\sc J.~A. Braatz}, {\sc S.~Muller}, {\sc J.~Ott}, and {\sc
  R.~Q. Mao},
\newblock {\em Astronomy {\&} Astrophysics} {\bf 500}, 725 (2009).

\bibitem{11MuBeGu.mp2me}
{\sc S.~Muller}, {\sc A.~Beelen}, {\sc M.~Gu{\'{e}}lin}, {\sc S.~Aalto}, {\sc
  J.~H. Black}, {\sc F.~Combes}, {\sc S.~J. Curran}, {\sc P.~Theule}, and {\sc
  S.~N. Longmore},
\newblock {\em Astronomy {\&} Astrophysics} {\bf 535}, A103 (2011).

\bibitem{11Ka.mp2me}
{\sc N.~Kanekar},
\newblock {\em The Astrophysical Journal} {\bf 728}, L12 (2011).

\bibitem{15KaUbMe.mp2me}
{\sc N.~Kanekar}, {\sc W.~Ubachs}, {\sc K.~M. Menten}, {\sc J.~Bagdonaite},
  {\sc A.~Brunthaler}, {\sc C.~Henkel}, {\sc S.~Muller}, {\sc H.~L. Bethlem},
  and {\sc M.~Dapr{\`{a}}},
\newblock {\em Monthly Notices of the Royal Astronomical Society: Letters} {\bf
  448}, L104 (2015).

\bibitem{13BaJaHe.mp2me}
{\sc J.~Bagdonaite}, {\sc P.~Jansen}, {\sc C.~Henkel}, {\sc H.~L. Bethlem},
  {\sc K.~M. Menten}, and {\sc W.~Ubachs},
\newblock {\em Science (New York, N.Y.)} {\bf 339}, 46 (2013).

\bibitem{06MeStIv.mp2me}
{\sc V.~V. Meshkov}, {\sc A.~V. Stolyarov}, {\sc A.~V. Ivanchik}, and {\sc
  D.~A. Varshalovich},
\newblock {\em Journal of Experimental and Theoretical Physics Letters} {\bf
  83}, 303 (2006).

\bibitem{05IvPeVa.mp2me}
{\sc A.~V. Ivanchik}, {\sc P.~Petitjean}, {\sc D.~A. Varshalovich}, {\sc
  B.~Aracil}, {\sc R.~Srianand}, {\sc H.~Chand}, {\sc C.~Ledoux}, and {\sc
  P.~Boiss{\'{e}}},
\newblock {\em Astronomy {\&} Astrophysics} {\bf 440}, 45 (2005).

\bibitem{16UbKoEi.mp2me}
{\sc W.~Ubachs}, {\sc J.~C.~J. Koelemeij}, {\sc K.~S.~E. Eikema}, and {\sc
  E.~J. Salumbides},
\newblock {\em Journal of Molecular Spectroscopy} {\bf 320}, 1 (2016).

\bibitem{13BaDaJa.mp2me}
{\sc J.~Bagdonaite}, {\sc M.~Dapr{\`{a}}}, {\sc P.~Jansen}, {\sc H.~L.
  Bethlem}, {\sc W.~Ubachs}, {\sc S.~Muller}, {\sc C.~Henkel}, and {\sc K.~M.
  Menten},
\newblock {\em Physical Review Letters} {\bf 111}, 231101 (2013).

\bibitem{09MaBuMu.mp2me}
{\sc A.~L. Malec}, {\sc R.~Buning}, {\sc M.~T. Murphy}, {\sc N.~Milutinovic},
  {\sc S.~L. Ellison}, {\sc J.~X. Prochaska}, {\sc L.~Kaper}, {\sc
  J.~Tumlinson}, {\sc R.~F. Carswell}, and {\sc W.~Ubachs},
\newblock {\em Mem. S.A.It} {\bf 80}, 882 (2009).

\bibitem{11Ch.mp2me}
{\sc T.~Chiba},
\newblock {\em Progress of Theoretical Physics} {\bf 126}, 993 (2011).

\bibitem{18Ub.mp2me}
{\sc W.~Ubachs},
\newblock {\em Space Science Reviews} {\bf 214}, 3 (2018).

\bibitem{12SaNiBa.mp2me}
{\sc E.~J. Salumbides}, {\sc M.~L. Niu}, {\sc J.~Bagdonaite}, {\sc
  N.~de~Oliveira}, {\sc D.~Joyeux}, {\sc L.~Nahon}, and {\sc W.~Ubachs},
\newblock {\em Physical Review A} {\bf 86}, 022510 (2012).

\bibitem{11JaKlXu.mp2me}
{\sc P.~Jansen}, {\sc I.~Kleiner}, {\sc L.~Xu}, {\sc W.~Ubachs}, and {\sc H.~L.
  Bethlem},
\newblock {\em Physical Review A} {\bf 84}, 062505 (2011).

\bibitem{08IvRoVi.mp2me}
{\sc T.~I. Ivanov}, {\sc M.~Roudjane}, {\sc M.~O. Vieitez}, {\sc C.~A.
  de~Lange}, {\sc W.-U.~L. Tchang-Brillet}, and {\sc W.~Ubachs},
\newblock {\em Physical Review Letters} {\bf 100}, 1 (2008).

\bibitem{12DeUbBe.mp2me}
{\sc A.~J. de~Nijs}, {\sc W.~Ubachs}, and {\sc H.~L. Bethlem},
\newblock {\em Physical Review A} {\bf 86}, 1 (2012).

\bibitem{11BeKoBo.mp2me}
{\sc K.~Beloy}, {\sc M.~G. Kozlov}, {\sc A.~Borschevsky}, {\sc A.~W. Hauser},
  {\sc V.~V. Flambaum}, and {\sc P.~Schwerdtfeger},
\newblock {\em Physical Review A} {\bf 83}, 062514 (2011).

\bibitem{07FlKo.mp2me}
{\sc V.~V. Flambaum} and {\sc M.~G. Kozlov},
\newblock {\em Physical Review Letters} {\bf 98}, 1 (2007).

\bibitem{15OwYuTh.mp2me}
{\sc A.~Owens}, {\sc S.~N. Yurchenko}, {\sc W.~Thiel}, and {\sc
  V.~{\v{S}}pirko},
\newblock {\em Monthly Notices of the Royal Astronomical Society} {\bf 450},
  3191 (2015).

\bibitem{04VeKuBe.mp2me}
{\sc J.~Veldhoven}, {\sc J.~K{\"{u}}pper}, {\sc H.~L. Bethlem}, {\sc
  B.~Sartakov}, {\sc A.~J.~A. Roij}, and {\sc G.~Meijer},
\newblock {\em The European Physical Journal D} {\bf 31}, 337 (2004).

\bibitem{16OwYuTh.mp2me}
{\sc A.~Owens}, {\sc S.~N. Yurchenko}, {\sc W.~Thiel}, and {\sc
  V.~{\v{S}}pirko},
\newblock {\em Physical Review A} {\bf 93}, 052506 (2016).

\bibitem{10KoLaLe.mp2me}
{\sc M.~G. Kozlov}, {\sc A.~V. Lapinov}, and {\sc S.~A. Levshakov},
\newblock {\em Journal of Physics B: Atomic, Molecular and Optical Physics}
  {\bf 43}, 074003 (2010).

\bibitem{11KoLe.mp2me}
{\sc M.~G. Kozlov} and {\sc S.~A. Levshakov},
\newblock {\em The Astrophysical Journal} {\bf 726}, 65 (2011).

\bibitem{11KoPoRe.mp2me}
{\sc M.~G. Kozlov}, {\sc S.~G. Porsev}, and {\sc D.~Reimers},
\newblock {\em Physical Review A} {\bf 83}, 1 (2011).

\bibitem{11Ko.mp2me}
{\sc M.~G. Kozlov},
\newblock {\em Physical Review A} {\bf 84}, 042120 (2011).

\bibitem{13JaXuKl.mp2me}
{\sc P.~Jansen}, {\sc L.~Xu}, {\sc I.~Kleiner}, {\sc H.~L. Bethlem}, and {\sc
  W.~Ubachs},
\newblock {\em Physical Review A} {\bf 87}, 1 (2013).

\bibitem{14Il.mp2me}
{\sc V.~V. Ilyushin},
\newblock {\em Journal of Molecular Spectroscopy} {\bf 300}, 86 (2014).

\bibitem{14ViKo.mp2me}
{\sc A.~V. Viatkina} and {\sc M.~G. Kozlov},
\newblock {\em Journal of Molecular Spectroscopy} {\bf 300}, 94 (2014).

\bibitem{18Mc.mp2me}
{\sc B.~A. McGuire},
\newblock {\em The Astrophysical Journal Supplement Series} {\bf 239}, 17
  (2018).

\bibitem{79HeHu.diatomic}
{\sc G.~Herzberg} and {\sc K.~P. Huber},
\newblock {Constants Of Diatomic Molecules},
\newblock in {\em Molecular Spectra and Molecular Structure}, p. 498, Van
  Nostrand Reinhold Company, 1979.

\bibitem{15LoYuTe.exomol}
{\sc L.~Lodi}, {\sc S.~N. Yurchenko}, and {\sc J.~Tennyson},
\newblock {\em Molecular Physics} {\bf 113}, 1998 (2015).

\bibitem{16McYuTe.exomol}
{\sc L.~K. McKemmish}, {\sc S.~N. Yurchenko}, and {\sc J.~Tennyson},
\newblock {\em Monthly Notices of the Royal Astronomical Society} {\bf 463},
  771 (2016).

\bibitem{10TeZi.abun}
{\sc E.~D. Tenenbaum} and {\sc L.~M. Ziurys},
\newblock {\em The Astrophysical Journal} {\bf 712}, L93 (2010).

\bibitem{16GeNeGo.abun}
{\sc M.~Gerin}, {\sc D.~A. Neufeld}, and {\sc J.~R. Goicoechea},
\newblock {\em Annual Review of Astronomy and Astrophysics} {\bf 54}, 181
  (2016).

\bibitem{13McWaMa.abun}
{\sc D.~McElroy}, {\sc C.~Walsh}, {\sc A.~J. Markwick}, {\sc M.~A. Cordiner},
  {\sc K.~Smith}, and {\sc T.~J. Millar},
\newblock {\em Astronomy and Astrophysics Supplement Series} {\bf 550}, 1
  (2013).

\bibitem{13Ti.abun}
{\sc A.~G. G.~M. Tielens},
\newblock {\em Reviews of Modern Physics} {\bf 85}, 1022 (2013).

\bibitem{16MuKaBl.abun}
{\sc S.~Muller}, {\sc K.~Kawaguchi}, {\sc J.~H. Black}, and {\sc T.~Amano},
\newblock {\em Astronomy {\&} Astrophysics} {\bf 589}, L5 (2016).

\bibitem{10AgCeGu.abun}
{\sc M.~Ag{\'{u}}ndez}, {\sc J.~Cernicharo}, {\sc M.~Gu{\'{e}}lin}, {\sc
  C.~Kahane}, {\sc E.~Roueff}, {\sc J.~K{\l}os}, {\sc F.~J. Aoiz}, {\sc
  F.~Lique}, {\sc N.~Marcelino}, {\sc J.~R. Goicoechea}, {\sc
  M.~Gonz{\'{a}}lez~Garc{\'{i}}a}, {\sc C.~A. Gottlieb}, {\sc M.~C. Mccarthy},
  and {\sc P.~Thaddeus},
\newblock {\em Astronomy {\&} Astrophysics} {\bf 517}, 2 (2010).

\bibitem{98CeCaWo.abun}
{\sc C.~Ceccarelli}, {\sc E.~Caux}, {\sc M.~Wolfire}, {\sc A.~Rudolph}, {\sc
  B.~Nisini}, {\sc P.~Saraceno}, and {\sc G.~J. White},
\newblock {\em Astronomy {\&} Astrophysics} , 17 (1998).

\bibitem{08MiHaTe.abun}
{\sc S.~N. Milam}, {\sc D.~T. Halfen}, {\sc E.~D. Tenenbaum}, {\sc A.~J.
  Apponi}, {\sc N.~J. Woolf}, and {\sc L.~M. Ziurys},
\newblock {\em The Astrophysical Journal} , 618 (2008).

\bibitem{03FuWaNa.abun}
{\sc R.~S. Furuya}, {\sc C.~M. Walmsley}, {\sc K.~Nakanishi}, {\sc P.~Schilke},
  and {\sc R.~Bachiller},
\newblock {\em Astronomy {\&} Astrophysics} {\bf 409}, 21 (2003).

\bibitem{13Ti.mp2me}
{\sc A.~G. G.~M. Tielens},
\newblock {\em Reviews of Modern Physics} {\bf 85}, 1021 (2013).

\bibitem{17PrSaCe.abun}
{\sc L.~V. Prieto}, {\sc C.~S{\'{a}}nchez~Contreras}, {\sc J.~Cernicharo}, {\sc
  M.~Ag{\'{u}}ndez}, {\sc G.~Quintana-Lacaci}, {\sc V.~Bujarrabal}, {\sc
  J.~Alcolea}, {\sc C.~Balan{\c{c}}a}, {\sc F.~Herpin}, {\sc K.~M. Menten}, and
  {\sc F.~Wyrowski},
\newblock {\em Astronomy {\&} Astrophysics} {\bf 597}, 25 (2017).

\bibitem{14CeBaAl.abun}
{\sc J.~Cernicharo}, {\sc S.~Bailleux}, {\sc E.~Alekseev}, {\sc A.~Fuente},
  {\sc E.~Roueff}, {\sc M.~Gerin}, {\sc B.~Tercero}, {\sc S.~P.
  Trevi{\~{n}}o-Morales}, {\sc N.~Marcelino}, {\sc R.~Bachiller}, and {\sc
  B.~Lefloch},
\newblock {\em The Astrophysical Journal} {\bf 795}, 40 (2014).

\bibitem{03MaMaMa.abun}
{\sc S.~Mart{\'{i}}n}, {\sc R.~Mauersberger}, {\sc J.~Mart{\'{i}}n-Pintado},
  {\sc S.~Garc{\'{i}}a-Burillo}, and {\sc C.~Henkel},
\newblock {\em Astronomy {\&} Astrophysics} {\bf 411}, 465 (2003).

\bibitem{18CeLeAg.abun}
{\sc J.~Cernicharo}, {\sc B.~Lefloch}, {\sc M.~Ag{\'{u}}ndez}, {\sc
  S.~Bailleux}, {\sc L.~Margul{\`{e}}s}, {\sc E.~Roueff}, {\sc R.~Bachiller},
  {\sc N.~Marcelino}, {\sc B.~Tercero}, {\sc C.~Vastel}, and {\sc E.~Caux},
\newblock {\em The Astrophysical Journal} {\bf 853}, L22 (2018).

\bibitem{03ScLeMe.abun}
{\sc P.~Schilke}, {\sc S.~Leurini}, {\sc K.~M. Menten}, and {\sc J.~Alcolea},
\newblock {\em Astronomy {\&} Astrophysics} {\bf 412}, L15 (2003).

\bibitem{16TeYuAl.exomol}
{\sc J.~Tennyson}, {\sc S.~N. Yurchenko}, {\sc A.~F. Al-Refaie}, {\sc E.~J.
  Barton}, {\sc K.~L. Chubb}, {\sc P.~A. Coles}, {\sc S.~Diamantopoulou}, {\sc
  M.~N. Gorman}, {\sc C.~Hill}, {\sc A.~Z. Lam}, {\sc L.~Lodi}, {\sc L.~K.
  McKemmish}, {\sc Y.~Na}, {\sc A.~Owens}, {\sc O.~L. Polyansky}, {\sc
  T.~Rivlin}, {\sc C.~Sousa-Silva}, {\sc D.~S. Underwood}, {\sc A.~Yachmenev},
  and {\sc E.~Zak},
\newblock {\em Journal of Molecular Spectroscopy} {\bf 327}, 73 (2016).

\bibitem{18YuBoGo.exomol}
{\sc S.~N. Yurchenko}, {\sc W.~Bond}, {\sc M.~N. Gorman}, {\sc L.~Lodi}, {\sc
  L.~K. McKemmish}, {\sc W.~Nunn}, {\sc R.~Shah}, and {\sc J.~Tennyson},
\newblock {\em Monthly Notices of the Royal Astronomical Society} {\bf 478},
  270 (2018).

\bibitem{17WoYuBe.exomol}
{\sc A.~Wong}, {\sc S.~N. Yurchenko}, {\sc P.~Bernath}, {\sc H.~S.~P.
  M{\"{u}}ller}, {\sc S.~McConkey}, and {\sc J.~Tennyson},
\newblock {\em Monthly Notices of the Royal Astronomical Society} {\bf 470},
  882 (2017).

\bibitem{17PrJaLo.exomol}
{\sc L.~Prajapat}, {\sc P.~Jagoda}, {\sc L.~Lodi}, {\sc M.~N. Gorman}, {\sc
  S.~N. Yurchenko}, and {\sc J.~Tennyson},
\newblock {\em Monthly Notices of the Royal Astronomical Society} {\bf 472},
  3648 (2017).

\bibitem{15PaYuTe.exomol}
{\sc A.~T. Patrascu}, {\sc S.~N. Yurchenko}, and {\sc J.~Tennyson},
\newblock {\em Monthly Notices of the Royal Astronomical Society} {\bf 449},
  3613 (2015).

\bibitem{18YuSiLo.exomol}
{\sc S.~N. Yurchenko}, {\sc F.~Sinden}, {\sc L.~Lodi}, {\sc C.~Hill}, {\sc
  M.~N. Gorman}, and {\sc J.~Tennyson},
\newblock {\em Monthly Notices of the Royal Astronomical Society} {\bf 473},
  5324 (2018).

\bibitem{18YuSzPy.exomol}
{\sc S.~N. Yurchenko}, {\sc I.~Szab{\'{o}}}, {\sc E.~Pyatenko}, and {\sc
  J.~Tennyson},
\newblock {\em Monthly Notices of the Royal Astronomical Society} {\bf 480},
  3397 (2018).

\bibitem{19McMaHo.exomol}
{\sc L.~K. McKemmish}, {\sc T.~Masseron}, {\sc H.~J. Hoeijmakers}, {\sc
  V.~P{\'{e}}rez-Mesa}, {\sc S.~L. Grimm}, {\sc S.~N. Yurchenko}, and {\sc
  J.~Tennyson},
\newblock {\em Monthly Notices of the Royal Astronomical Society} {\bf 488},
  2836 (2019).

\bibitem{07De.mp2me}
{\sc T.~Dent},
\newblock {\em Journal of Cosmology and Astroparticle Physics} {\bf 2007}, 013
  (2007).

\bibitem{16YuLoTe.mp2me}
{\sc S.~N. Yurchenko}, {\sc L.~Lodi}, {\sc J.~Tennyson}, and {\sc A.~V.
  Stolyarov},
\newblock {\em Computer Physics Communications} {\bf 202}, 262 (2016).

\bibitem{17TeYu.exomol}
{\sc J.~Tennyson} and {\sc S.~N. Yurchenko},
\newblock {\em International Journal of Quantum Chemistry} {\bf 117}, 92
  (2017).

\bibitem{14BrRaWe.mp2me}
{\sc J.~S.~A. Brooke}, {\sc R.~S. Ram}, {\sc C.~M. Western}, {\sc G.~Li}, {\sc
  D.~W. Schwenke}, and {\sc P.~F. Bernath},
\newblock {\em The Astrophysical Journal Supplement Series} {\bf 210}, 23
  (2014).

\bibitem{14RaBrWe.mp2me}
{\sc R.~S. Ram}, {\sc J.~S.~A. Brooke}, {\sc C.~M. Western}, and {\sc P.~F.
  Bernath},
\newblock {\em Journal of Quantitative Spectroscopy and Radiative Transfer}
  {\bf 138}, 107 (2014).

\bibitem{13TrHeTo.mp2me}
{\sc S.~Truppe}, {\sc R.~J. Hendricks}, {\sc S.~K. Tokunaga}, {\sc H.~J.
  Lewandowski}, {\sc M.~G. Kozlov}, {\sc C.~Henkel}, {\sc E.~A. Hinds}, and
  {\sc M.~R. Tarbutt},
\newblock {\em Nature Communications} {\bf 4}, 2600 (2013).

\bibitem{DVR3D}
{\sc J.~Tennyson}, {\sc M.~A. Kostin}, {\sc P.~Barletta}, {\sc G.~J. Harris},
  {\sc O.~L. Polyansky}, {\sc J.~Ramanlal}, and {\sc N.~F. Zobov},
\newblock {\em Computer Phys. Comm.} {\bf 163}, 85 (2004).

\bibitem{TROVE}
{\sc S.~N. Yurchenko}, {\sc W.~Thiel}, and {\sc P.~Jensen},
\newblock  {\bf 245}, 126 (2007).

\bibitem{15PaBoFl.mp2me}
{\sc L.~F. Pa{\v{s}}teka}, {\sc A.~Borschevsky}, {\sc V.~V. Flambaum}, and {\sc
  P.~Schwerdtfeger},
\newblock {\em Physical Review A} {\bf 92}, 1 (2015).

\end{thebibliography}
\end{document}